\documentclass[12pt,preprint]{aastex} 

\def\etal{et\thinspace al.\ }                               

\shortauthors{Gonz\'alez Delgado et al.}
\shorttitle{HST/WFPC2 imaging of LLAGNs}

\begin{document}

\title{HST/WFPC2 imaging of the circumnuclear structure of LLAGNs. I Data and nuclear morphology\footnote{Based on observations with NASA/ESA {\it Hubble Space Telescope} obtained at the Space Telescope Science Institute, which is operated by the Association of Universities for Research in Astronomy, Inc., under NASA contract NAS 5-2655. }}

\author{
Rosa M. Gonz\'alez Delgado\altaffilmark{1},
Enrique P\'erez\altaffilmark{1},
Roberto Cid Fernandes\altaffilmark{2},
Henrique Schmitt\altaffilmark{3,}}

\affil{(1) \em Instituto de Astrof\'{\i}sica de Andaluc\'{\i}a (CSIC), P.O. Box 3004, 18080 Granada, Spain (rosa@iaa.es; eperez@iaa.es)}
\affil{(2) \em Depto. de F\'\i sica-CFM, Universidade Federal de Santa 
Catarina, C.P. 476, 88040-900, Florian\'opolis, 
SC, Brazil (cid@astro.ufsc.br)}
\affil{(3) \em Remote Sensing Division, Naval Research Laboratory, Code 7210, 4555 Overlook Avenue, Washington, DC\,20375 (hschmitt@ccs.nrl.navy.mil)}
\affil{(4) \em Interferometrics, Inc., 13454 Sunrise Valley Drive, Suite 240, Herndon, VA\,20171}

\begin{abstract}

In several studies of Low Luminosity Active Galactic Nuclei (LLAGNs),
we have characterized the properties of the stellar populations in
LINERs and LINER/HII Transition Objects (TOs).  We have found a
numerous class of galactic nuclei which stand out because of their
conspicuous 0.1--1 Gyr populations. These nuclei were called
''Young-TOs'' since they all have TO-like emission line ratios. To
advance our knowledge of the nature of the central source in LLAGNs
and its relation with stellar clusters, we are carrying out several
imaging projects with the Hubble Space Telescope (HST) at near-UV,
optical and near-IR wavelengths.  In this paper, we present the first
results obtained with observations of the central regions of 57 LLAGNs
imaged with the WFPC2 through any of the V (F555W, F547M, F614W) and I
(F791W, F814W) filters that are available in the HST archive. The
sample contains 34$\%$ of the LINERs and 36$\%$ of the TOs in the
Palomar sample.  The mean spatial resolution of these images is 10
pc. With these data we have built an atlas that includes structural
maps for all the galaxies, useful to identify compact nuclear
sources and, additionally, to characterize the circumnuclear
environment of LLAGNs, determining the frequency of dust and its
morphology.  The main results obtained are: 1) We have not found any
correlation between the presence of nuclear compact sources and
emission-line type.  Thus, nucleated LINERs are as frequent as
nucleated TOs.  2) The nuclei of "Young-TOs" are brighter than the
nuclei of "Old-TOs" and LINERs. These results confirm our previous
results that Young-TOs are separated from other LLAGNs classes in
terms of their central stellar population properties and brightness.
3) Circumnuclear dust is detected in 88$\%$ of the LLAGNs, being
almost ubiquitous in TOs.  4) The dust morphology is complex and
varied, from nuclear spiral lanes to chaotic filaments and nuclear
disk-like structures. Chaotic filaments are as frequent as dust
spirals; but nuclear disks are mainly seen in LINERs.  These results
suggest an evolutionary sequence of the dust in LLAGNs, LINERs
being the more evolved systems and Young-TOs the youngest.
\end{abstract}


\section{Introduction}
\label{sec:Introduction}


Low-luminosity active galactic nuclei (LLAGNs) comprise 30\%\ of all
bright galaxies (B$\leq$12.5) and are the most common type of AGN (Ho,
Filippenko \& Sargent 1997a, hereafter HFS97). These include LINERs,
and transition-type objects (TOs, also called weak-[OI] LINERs). These
two types of LLAGNs have similar emission line ratios in
[OIII]/H$\beta$, [NII]/H$\alpha$, and [SII]/H$\alpha$, but
[OI]/H$\alpha$ is lower in TOs than in LINERs. LLAGNs constitute a
rather mixed class and different mechanisms have been proposed to
explain the origin of the nuclear activity, including shocks, and
photoionization by a non-stellar source, by hot stars or by
intermediate age stars (e.g.  Ferland \& Netzer 1983; Filippenko \&
Terlevich 1992; Binnette et al. 1994; Taniguchi, Shioya \& Murayama
2000).  Because we do not know yet what powers them and how they are
related to the Seyfert phenomenon, LLAGNs have been at the forefront
of AGN research since they were first  systematically studied by
Heckman (1980).  Are they all truly ``dwarf'' Seyfert nuclei powered
by accretion onto nearly dormant supermassive black holes (BH), or can
some of them be explained at least partly in terms of stellar
processes? If LLAGNs were powered by a BH, they would represent the
low end of the AGN luminosity function in the local universe and would
also establish a lower limit to the fraction of galaxies containing
massive BHs in their centers. If, on the contrary, LLAGNs were powered
by nuclear stellar clusters, their presence would play an important
role in the evolution of galaxy nuclei. Therefore, it is fundamental
to unveil the nature of the central source in LLAGNs.

There is clear evidence that at least some LINERs harbor a bona fide
AGN, and they may be considered the faint end of the luminosity function
of Seyfert galaxies. It has been found that about 20\%\ of the nearby
LINERs have a weak broad H$\alpha$ emission component similar to 
those found in type 1 Seyferts (Ho, Filippenko \& Sargent 1997b). In a
few of these LINERs the H$\alpha$ line shows a double peak component
(e.g.  Storchi-Bergmann et al. 1997; Shields et al. 2000; Ho et
al. 2000) suggesting that they are powered by an accreting black hole
(BH). It has also been found that X-ray emission in LINERs has a
nonthermal origin associated with an AGN (e.g. Terashima, Ho \& Ptak
2000).  Recent works based on higher spatial resolution data taken
with Chandra show that only in half
of the LLAGNs observed the
X-ray emission is associated with compact
nuclear cores (Satyapal et
al. 2004; Dudik et al. 2006; Gonz\'alez-Mart\'\i n et
al. 2006). 
This is consistent with VLA radio observations, which
detect unresolved radio cores in half of the LINERs 
 (Nagar et
al. 2000, 2002). Finally, a monitoring study at near-UV wavelengths by
Maoz et al. (2005)
 finds UV variability in a significant
fraction of the 17 LLAGNs observed. The variation of the UV fluxes may
be interpreted as the manifestation of low rate or low radiative
efficiency accretion onto a supermassive BH.

On the other hand, detection of stellar wind absorption lines in the
ultraviolet spectra of some TOs (Maoz et al. 1998; Colina et al. 2002)
has proven unequivocally the presence of young stellar clusters in the
nuclear region. Additional evidence comes from optical studies, in
which we have focused on the study of the stellar population in the
nuclei and circumnuclear region of LLAGNs to establish the role of
stellar processes in their phenomenology. Ground-based
(Cid Fernandes et al.\ 2004; Cid Fernandes et al.\ 2005) and HST+STIS
spectra (Gonz\'alez Delgado et al.\ 2004) have shown that the
contribution of an intermediate age stellar population is significant
in a sizable fraction of the TO population.  These studies
identified a class of objects, called ``Young-TOs'', which are clearly
separated from LINERs in terms of the properties and spatial
distribution of the stellar populations. They have stronger stellar
population gradients, a luminous intermediate age stellar population
concentrated toward the nucleus ($\sim$100~pc) and much larger amounts
of extinction than LINERs. These objects, which underwent a powerful
star formation event $\sim$ 1 Gyr ago, could correspond to
post-Starburst nuclei or to evolved counterparts of the Seyfert 2 with
a composite nucleus, characterized too by harboring nuclear starbursts
(Gonz\'alez Delgado et al. 2001; Cid Fernandes et al. 2001, 2004). HST
imaging of the nuclei of these Seyfert 2 galaxies shows that the UV
emission is resolved into stellar clusters (Gonz\'alez Delgado et al.
1998) that are similar to those detected in starburst galaxies (Meurer
et al.\ 1995).

Nuclear stellar clusters are a common phenomenon in spirals, having
been detected in 50-70\% of these sources (Carollo et al.  1998, 2002;
Boeker et al. 2002, 2004). Therefore stellar clusters are a natural
consequence of the star formation processes in the central region of
spirals. On the other hand, evidence has been accumulating during the
past few years about the ubiquity of BH in the nuclei of
galaxies. Furthermore, the tight correlation of the BH mass and
stellar velocity dispersion (Ferrarese \& Merrit 2000; Gebhardt et
al.\ 2000) implies that the creation and evolution of a BH is
intimately connected to that of the galaxy bulge. Recently, in a
HST survey in the Virgo Cluster, C\^ot\'e et al. (2006) have detected
compact sources in a comparable fraction of elliptical galaxies.
These compact stellar clusters, referred to as nuclei by the
authors (see also Ferrarese et al 2006a), have masses that scale
directly to the galaxy mass, in the same way as do the BH masses in
high luminosity galaxies (Ferrarese et al. 2006b). Therefore, a
natural consequence of the physical processes that formed present-day
galaxies should be the creation of a compact massive object in the
nucleus, either a BH and/or a massive stellar cluster.

To determine the nature of the nuclear source of active galaxies and
their evolution we are carrying out several projects with HST+ACS
imaging at the near-UV wavelengths a sample of Seyferts (ID. 9379,
PI. Schmitt, Mu\~noz-Mar\'\i n et al. 2007) and LLAGNs
(ID. 10548, PI.  Gonz\'alez Delgado). These observations are
complemented with WFPC2 optical data retrieved from the HST archive. 
The high angular resolution provided
by HST is crucial to determine the physical properties of the nuclei,
the central structure of these galaxies, as well as to study the
circumnuclear environment of AGNs. The main goals of these studies are
to determine the frequency of nuclear and circumnuclear stellar
clusters in AGNs, and whether they are more common in Seyferts, TOs or
LINERs; to characterize the intrinsic properties of these clusters and
to study whether there is evolution from Seyferts to TOs and LINERs.
In addition, the frequency of dust and its morphology can also provide
relevant information about the origin of nuclear activity. Dust is
a valuable probe of the presence of cold interstellar gas in galaxies,
and it is very sensitive to the perturbations that drive the gas
toward the center and feed the AGN. Here, we present the initial
results obtained for LLAGNs based on archival visible and red images
obtained with the WFPC2.  The paper is organized as follows: section 2
presents the sample selection, and 3 the characteristics of the
observations. Sections 4, 5 and 6 describe the imaging atlas, the dust
morphology and the central properties of the galaxies. Finally, the
summary and conclusions are presented in section 7.


\section{Galaxy Sample}

\label{sec:Sample}

The objects selected for this study are drawn entirely from the HFS97
catalog as it comprises the most complete and homogeneous survey of
LINERs and TOs available for the local universe. Of the 160 galaxies
classified as LLAGNs, we have already studied the cirmnuclear stellar
population for about half of the HFS97 catalog (Cid Fernandes et al.\
2004; Gonz\'alez Delgado et al.\ 2004; Cid Fernandes et al.\ 2005;
hereafter Papers I, II and III, respectively). These galaxies
constitute the raw sample for this study. We have found suitable WFPC2
archival images for 57 of them, i.e.  36\% of the original HFS97
LLAGNs sample. 

Following the HFS97 emission-line classification, our subset contains
32 LINERs ([OI]/H$\alpha > 0.17$ ) and 25 TOs ([OI]/H$\alpha \le
0.17$), i.e., 34 and 38\% of whole HFS97 LLAGN sample, respectively.
In Paper I we proposed a slight modification of this criterion,
setting the LINER/TO dividing line at [OI]/H$\alpha = 0.25$. According
to this definition, to be used throughout the rest of this paper, our
sample includes 36\% of the TOs and 34\% of the LINERs in the HFS97
catalog. Figure 1 shows the morphological type, distance and
[OI]/H$\alpha$ distributions of our sample as compared with the full
HFS97 sample. The objects studied here generally follow the same
distributions that the HFS97 sample. In our sample, the mean distance
is 17 Mpc for the weak-[OI] ([OI]/H$\alpha <= 0.25$) and 30 Mpc for
the strong-[OI] ([OI]/H$\alpha > 0.25$) sources. The
corresponding values in the whole HFS97 LLAGNs sample are 23 Mpc and
28 Mpc. The median morphological types in both samples are S0 and Sab
for strong and weak-[OI], respectively.

Table 1 lists the main properties for the galaxies selected, such as
emission spectral type, Hubble type, distance, and stellar population
classification. The emission-line class, Hubble type and
distance are from HFS97. These authors adopted the Hubble type
from the RC3 (de Vaucouleurs et al 1991), and the distance from Tully
(1988) for galaxies closer than 40 Mpc. For farther galaxies
they derive distances simply from their radial velocities and
$H_0 = 75$ km$\,$s$^{-1}$ Mpc$^{-1}$. The properties related to the
stellar population are from papers I and II. Following these
papers, systems are classified as Young (Y) or Old (O) based on the
detection of intermediate age stellar population features and the
value of the equivalent width of the CaII K absorption band. Y nuclei
always have $W_K\le 15$ \AA. Most of the O-nuclei have $W_K > 15$ \AA,
however, there are four LINERs, NGC 3998, NGC 4143, NGC 4203, and NGC
4450, that have old nuclear stellar populations but the CaII lines are
diluted due to the AGN contribution, and thus they violate the
$W_K > 15$ \AA\ rule for O systems.

In terms of combined emission ([OI]/H$\alpha$) and stellar population
classifications, the current subsample comprises 17 Young-TOs,
20 Old-TOs, 18 Old-LINERs and 2 Young-LINERs.  As we have found
previously, $\sim$ half of the TOs have young ($\leq$ 1 Gyr)
populations, and almost no LINER has young stars.


\section{Observations and data processing}

The data collected for our subset of LLAGNs were all obtained with the WFPC2
on board of HST at optical wavelengths through any of these filters:
F555W, F547W, F606W, F791W and F814W. The first three filters are a proxy
for the visual V band, and the last two filters for the red broad I band.
Table 2 lists for each galaxy the filters, the exposure times, and the
image spatial scale of the observations, and the proposal identification
number. Most of the galaxies were observed through one of the filters close
to the V band, while three galaxies (NGC 4435, NGC 4438 and NGC 7331) have
been observed only with the I band filters. Most of the galaxies (86\%)
were imaged with their centers in the high-resolution Planetary Camera (PC)
CCD of the WFPC2, thus with a spatial sampling of 0.0456 arcsec/pixel, and
a field of view of 36$\times$36 arcsec. The remaining 8 galaxies were
observed with a sampling of 0.1 arcsec/pixel and a field of view of
1.3$\times$1.3 arcmin. Observations in both I and V equivalent filters are
available for 60\% of the sample, and for 40\% of the galaxies we have a
color image (V-I) with the high spatial sampling of 0.0456 arcsec/pixel.

The images were all processed by the standard WFPC2 pipeline developed
at the STScI that corrects for flat-field and bias subtraction. When
multiple exposures are available for the same target, the frames were
combined and cosmic-ray cleaned in a single step using the IRAF/STSDAS
task {\tt crrej}, otherwise cosmic rays were removed using the
task {\tt cosmicrays} in the IRAF\footnote{IRAF is distributed by the
National Optical Astronomical Observatory, which is operated by AURA,
Inc., under contract to the NSF.} package. For each frame, the
cosmic-ray events were defined as pixels deviating by more than 5
sigma above the background and they were replaced by the average of
several neighboring pixels. We have checked that the nucleus was not
compromised by the 
cosmic ray rejection algorithm, i.e. removing
flux that really corresponds to the nucleus;
this could happen if the
central surface brightness increases very rapidly in the central 2-3
pixels.
To check this in the case of a single exposure, we have
compared the number counts
and the surface brightness profiles
through the maximum number count pixel before and 
after the cosmic
ray algorithm was applied. If the image results from the combination
of several 
frames, we compare the surface brightness of the
individual frames between them and with 
the resulting combined
image.

The sky background of the images was estimated measuring in the outer part
of the mosaiced WFPC2 image. We did not
attempt to subtract it because most of the galaxies are extended well
beyond the WFPC2 field of view; thus, this measurement provides a value
higher than the actual sky background. We checked that for most of
the galaxies these measurements, even though higher than the expected sky
background, are low enough to have a negligible effect on the measurements
corresponding at least to the central 5 arcsec derived with no sky
subtraction.

Photometric calibration was obtained using the zero point
provided by the PHOTFLAMB key-word in the image header. 
PHOTFLAMB is
the flux of a source with constant flux per unit wavelength which
produces a count rate of 1 DN/s.
For a given filter, the PHOTFLAMB
values for most of the data are quite similar, and the differences
between them 
is less than 1$\%$. However, a few images have been
observed with a gain equal to 15 e$^-$/DN, instead of 7 e$^-$/DN.
For
these images, the PHOTFLAMB is almost a factor 2 larger than for
the
 rest of the galaxies that were observed with gain equal to 7
e$^-$/DN. We have checked that the PHOTFLAMB values 
 are in
agreement with the value of this parameter provided by the package
synphot. We use the expression $-2.5\times \log$ (counts/s
$\times$ PHOTFLAMB)$ - 21.1$ to calculate the magnitudes in the STMAG
system.

For the filters used in this work, this method provides
zero points that are equivalent to the zero points
calculated as
Z$_{STMAG} +2.5 \times \log {\rm GR}_i + 0.1$ proposed by
Holtzman et al (1995).  The 
 Z$_{STMAG}$ is the synthetic zero
points tabulated in table 9 of Holtzman et al (1995); GR$_i$ is the
gain ratio (so, GR$_i$= 2 for gain=7 e$^-$/DN and GR$_i$= 1 for gain=
15 e$^-$/DN).
 Note that the Holtzman et al (1995) zero points are
derived using apertures of 0.5 arcsec, but the PHOTFLAMB are referred
to counts measured with an infinite aperture.  Because the infinite
aperture is defined as having 1.096 times the flux in an aperture with
0.5 arcsec, we need to add 0.1 mag to the Z$_{STMAG}$ to correct to
infinite aperture.

Correction for Galactic extinction was performed using the E(B$-$V) values
given in the NED\footnote{The NASA/IPAC Extragalactic Database (NED)
is operated by the Jet Propulsion Laboratory, California Institute of
Technology, under contract with the National Aeronautics and Space
Administration.}  following Schlegel et al (1998), and 
with the Cardelli et al. (1989) reddening law with 
$R_V = 3.1$.


\section{The atlas, colors and the structure maps}

An atlas of our sample is presented in Figure 2. For each galaxy, we
show the V equivalent image in three scales, the full PC frame (when
available, or the central $\pm50$ arcsec of the mosaiced image when
the nucleus is not in the PC chip), a zoom into the central $\pm 4$
arcsec, and the central $\pm 10$ arcsec of the median filtered
contrast image. The galaxies display a significant variety of
morphologies. A brief description of the morphological features are
given in Table 3. The description indicates the dust classification,
the incidence of nuclear sources and other features. The criteria for
the dust classification and for determining whether a galaxy has a
nuclear source are further explained in sections 5 and 6,
respectively.

The V-I color image for 32 galaxies is also available, and in 23 of
these galaxies the color map is sampled with  0.0456 arcsec/pixel. 
Color images are displayed in Fin toigure~3.

There are several ways to map the circumnuclear dust in galaxies, each
with its own advantages and caveats. In early type galaxies
in which the central light distribution is well reproduced by
elliptical isophotes, the structure map is obtained simply subtracting
the model fit from the original image. The resulting image enhances
the small deviation of the dust fine structural features from the
isophotal model. This technique has been applied very often; some
recent results obtained with HST images are, for example, those of
Lauer et al. (2005). However, this technique can not be applied to all
types of galaxies, since many of them show central isophotes that
clearly deviate from an elliptical model. We were able to obtain
suitable isophotal fits for some of the galaxies in the sample. The
results of these models will be discussed in a forthcoming paper
(Gonz\'alez Delgado et al. 2007, in preparation).

Color images can also map the circumnuclear dust; for example,
Ferrarese et al (2006a) use g-z color for galaxies in the Virgo
cluster. Ideally, the color image must be obtained with two bands widely 
separated in wavelength, and with the same spatial resolution. 
This technique have been succesfully used to determine the amount of internal reddening in early type galaxies, where the stellar population does not change by a significant amount, so all the color variations can be attributed to dust. However, this assumption no longer applies for late type galaxies, which can show significant stellar population variations in the nucleus. Since many of our galaxies are spirals and we have two band frames (V and I) are available for only $\sim$60\% of the sample, we do not attempt to estimate the internal reddening using this technique.

To be able to obtain a structural map for all the galaxies, we have adopted
here the unsharp masking technique. The unsharp masked image is obtained by
dividing the original frame by a smoothed version obtained with a
31$\times$31 median boxcar kernel. The structural maps are displayed,
together with the original frames, in Figure 2. The dust structures are
shown as dark regions in the maps; bright regions are locations of emission
lines, if the filter includes emission lines, or enhanced stellar light.
This technique gives similar results to that proposed by Pogge \& Martini
(2002) which is based on the Richardson-Lucy image restoration process
(Richardson 1972; Lucy 1974), and it has been recently investigated by Simoes 
Lopes et al. (2007).

A comparison between the similarities and differences of the structural
maps obtained with the unsharp masking technique, the residual image
resulting by subtracting an isophotal model, and the color image are
displayed in Figure 4 for the galaxy NGC 3489. The main difference arises
from the sharp increases of surface brightness of the central light. 
Note, however, these unsharp images are used mainly for the dust
classification and not to find out whether a galaxy has a nuclear source.


\section{The circumnuclear dust}

HST observations of the central regions of galaxies have provided
significant results with respect to the detection of dust, its
morphology and its role in AGN.  In fact, dust provides indirect
evidence of the presence of cold interstellar gas and the mechanisms
that drive the gas to the nucleus and fuel the AGN. Circumnuclear
dust is common in Seyfert galaxies (e.g. Malkan, Gorjian \& Tran 1998;
Martini \& Pogge 1999; Regan \& Mulchaey 1999; Pogge \& Martini 2002).
Nuclear dust spirals are detected in many Seyfert galaxies. It has
been argued that they trace the paths followed by the gas in its way
to feed the AGN. A more recent study by Martini et al. (2003) has
confirmed the ubiquity of dust at the center of Seyfert galaxies, with
only 3\% of these galaxies not having nuclear dust structures
detected. These authors also find other dust morphologies in Seyfert
galaxies, which are less structured and more chaotic than the nuclear
spirals. Circumnuclear dust is also a common feature in radio galaxies
(Verdoes Kleijn \& Zeeuw 2005) and early type galaxies (Tran et
al. 2001; Lauer et al. 2005; Ferrarese et al 2006a).  In particular,
Lauer et al. (2005) have found that half of the galaxies in their
sample of early types have nuclear dust features, but this fraction
rises to 90\% when they consider only the galaxies with emission lines
(presumably associated with nuclear activity).  This result has been
recently confirmed by Simoes Lopes et al. (2007), who found that all
the 34 early-type AGN hosts in their sample have circumnuclear dust,
but dust is detected only in 26$\%$ of the early type non-active
galaxies.

We have found that circumnuclear dust is also common in LLAGNs, since
88\% of the galaxies in our sample display dust features. The
morphology is quite diverse, from nuclear disks, to filaments and
lanes chaotically distributed, to well organized nuclear spiral
arms. Here the morphology is determined by eye inspections,
independently carried out by three of the authors, following the
classification criteria established by Martini et al. (2003) and Lauer
el al. (2005). These criteria are complementary.  Lauer et
al. (2005) distinguish between nuclear rings and disks from other
structures that they divide into spiral and chaotic. The Martini et
al. (2003) classification, which is in some aspects an extension of
the Lauer et al. (2005) scheme, divides the spiral class into
three sub-classes: grand design nuclear spirals, tightly wound
spirals, and loosely wound nuclear spirals.  Note that these works
found significantly different results, that justify their
complementary dust classification.  For instance,
Martini et al. (2003) do not find any nuclear
disks in Seyfert galaxies, while in Lauer et al. (2005) the detection
of nuclear spirals in early type galaxies is rare. However, here we
detect a significant fraction of nuclear dust disks and different
types of spiral dust structures. Thus, we adopt the following criteria
for the classification:

\begin{itemize}

\item{{\it Grand Design Nuclear Spiral (GD)}. Galaxies belonging to this
class have two symmetric dust spiral arms. They are similar to the grand
design spirals but in spatial scales of less than a kiloparcsec. Only two (NGC 841, NGC 3705)
 of the LLAGNs of our sample belong to this class (4\% of the sample). }

\item{{\it Tightly Wound Nuclear Spiral (TW)}. Galaxies with
nuclear dust spirals with small pitch angles. They lack the symmetry of
the grand design class, and represent 19\% of the sample. 
One example of this class is NGC 4736.}

\item{{\it Loosely Wound Nuclear Spiral (LW)}. These have a coherent
nuclear spiral structure but with large pitch angle. One example is
NGC 6951. This class is $\sim$ as common as the TW class, with
14\% of our sample belonging to it.}

\item{{\it Chaotic Nuclear Spiral (CS)}. These galaxies show fragments of
dust arcs that suggest a spiral structure. Galaxies belonging to this class
may represent objects in a transition phase between the spiral and the
chaotic class. Examples of this class are NGC 3368 and NGC 3169, with
17\% of the LLAGNs in this class.  }

\item{{\it Chaotic Circumnuclear Dust (C)}. These galaxies
clearly show circumnuclear dust but they cannot be classified as
spiral structures. In many of these galaxies the dust appears as
filaments and lanes. 25\% of the LLAGNs are in this class, one example
is NGC 2685.  }

\item{{\it Disks and nuclear rings (D)}. They are circular and
axisymmetric structures. NGC 315 and NGC 2787 are examples of galaxies
belonging to this class. They represent 9\% of the sample.}

\item{{\it No dust structure (N)}. Galaxies with no dust in the
circumnuclear regions. NGC 3998 is one example; 12\% of LLAGNs do not
show any circumnuclear dust.}

\end{itemize}

Table 3 summarizes the description of the dust morphology
obtained, and Table 4 compares the frequency of dust morphologies for
the LLAGNs with the results for Seyferts and non-active galaxies. The
most remarkable result is that dust is almost ubiquitous in
LLAGNs. Only 12\% of the galaxies of our sample do not display dust
features. Dust is as frequent as in Seyferts (Martini et al. 2003), but
the distribution of the dust morphology is different. As mentioned
above, most of the Seyferts have dust spirals or dust filaments and
arcs chaotically distributed; the dust disk morphology and non-dusty
class are quite rare, while LLAGNs display dust disk
morphology.

Figure 5 shows the histogram of
the circumnuclear dust of LLAGNs in the seven dust-classes discussed.
Well organized nuclear dust spirals (GD+TW+LW) are as frequent as the
chaotic dust features (CS+C), and they are both more often detected
than the dust disk-like morphology. Figure 5 also displays the
distribution of the circumnuclear dust classes for strong- and
weak-[OI] LLAGNs, ie., LINERs and TOs. A significant differential
result emerges. The fraction of objects with dusty disks or no dust is
higher in LINERs than TOs, while the fraction of chaotic dust features
is equal in both types.  An even larger systematic difference appears dividing
the sample into the Young and Old stellar population categories. No
Young LLAGN falls in either the no-dust or dusty-disk classes. Old
systems, on the other hand, span the full menu of dust morphologies (Figure 6).

The spectral analysis performed in Papers I--III
strongly suggests that Young-TOs are dustier than other LLAGN.
A qualitative inspection of the images and structure maps in our atlas
reveals that dust features in these objects are indeed very strong,
which provides direct confirmation that Young-TOs stand out among
LLAGN in terms of their dust content.

Besides the morphology, we have also evaluated the dust concentration.
We have classified the LLAGNs in three classes according with the
detection of dust in the inner part, in the outer part or
throughout. Thus, the dust is highly concentrated if it is detected
only in the central region within a radius $\leq$ 100--200 pc,
which at the mean distance of the objects corresponds to 1--2 arcsec
radius. Dust concentration is graded by {\it in}, {\it out} or {\it
inout} if the dust is detected at shorter distances (radius $\leq$
100--200 pc), only at larger distances or at both,
respectively. We find that in most cases dust is distributed inside
and outside of the central 100--200 pc radius. We have not found any
difference in the dust concentration distribution in weak-[OI] and
strong-[OI] LLAGNs. Dust located only in the inner part is equally
frequent in TOs than in LINERs. However, the fraction of LLAGNs that
have the dust located only in the inner part is larger in
Old-systems (Old-TOs and LINERs) than in Young-systems (mainly
Young-TOs). For most of the Young-TOs the dust is located throughout.

Lauer et al. (2005) and Ferrarese et al. (2006a) have proposed that
the different dust morphologies in early type galaxies indicate a
dust-settling evolutionary sequence.  According to these
authors, dust would be initially distributed in chaotic and non-ring
structures, then it would sink to the center forming a disk or a ring,
and finally be destroyed. The results obtained here suggest too an
evolutionary sequence of the dust in these galaxies. If LLAGNs are
part of this sequence, LINERs are more evolved systems than TOs
because the fraction of objects with disks or non-dust detection is
significantly higher in the former type. This sequence is also in
agreement with the difference of nuclear stellar population results
found for these two types of LLAGNs (papers I and II). We found that
strong-[OI] LLAGNs are mainly characterized by an old stellar
population at the nucleus, but half of the weak-[OI] LLAGNs are
dominated by an intermediate stellar population (our
Young-TO class). In these works we suggest too an
evolutionary sequence from Young-TOs to Old-LINERs.

Other interesting differential results are also found dividing our
LLAGN sample in early (S0+E) and late (spirals) Hubble-types.  Dust is
almost ubiquitous in spirals, but not detected in 24$\%$ of the early
type LLAGN hosts.
This result is not totally in agreement with those
obtained by Simoes Lopes et al (2007), since they found that all the
early type AGN hosts of their sample have circumnuclear dust.
Additionally, we have found that the dust disk morphology is only
present in early type LLAGNs (see Table 4).


\section{Properties of the galaxy center}

\subsection{Central magnitude and surface brightness}

To determine the central magnitude we have performed circular aperture
photometry using the task {\tt phot} in the IRAF package SYNPHOT.  
Although the better way to measure the magnitudes of the galaxies would be by fitting ellipses to the images, this could only be done for $\sim$40\% of the sample. Since the nuclear regions of most galaxies are disturbed by dust or have a complicated stellar population distribution, they cannot be properly fitted by ellipses. Given these limitations, in order to obtain a consistent set of measurements for all galaxies we decided to use circular apertures.
The apertures are centered at the maximum brightness, which is determined
as the centroid of the central 5{\tt"}$\times$5{\tt"}. This criterium
is not applied for galaxies that have the center totally
obscured by dust lanes. In these cases, the centroid is calculated
with a larger number of central pixels, or it is taken as the center
of external isophotes which are not much affected by the dust
lanes. The integrated count rates are converted to ST magnitudes using
the PHOTFLAMB parameter of the image header as explained in section
3.  These central magnitudes are corrected for Galactic extinction
as described in Section 3.  Table 5 lists the central
magnitudes measured with apertures of 0.2, 0.5 and 1 arcsec radius in
in the filters F547M, F555W, F606W and F814W.
These quantities have not been corrected for sky background; the
reason is that most galaxies in the sample extend beyond the WFPC2 FOV
and thus the sky background can not be determined properly. However
these central quantities are not affected significantly by the sky
background. In fact, for those galaxies that are smaller than the
WFPC2 FOV, we have checked that the difference between the central
magnitude corrected and uncorrected for background is much less than
0.01 magnitude.  Note that these magnitudes correspond to integrated values, including the contribution from the stars and an AGN component when present, which have not been corrected for internal dust obscuration.

Additionally, the surface brightness ($\mu$) at several distances
(0.2, 0.5 and 1 arcsec) from the centroid is also calculated for each
galaxy and listed in Table 5. The approach is to extract
the mean number of counts in a circular annulus and then convert it to
magnitudes per arcsec$^2$. The measurements are performed in a
sequence of circular apertures of radii increasing in 1 pixel. This
method allows us to derive a surface brightness profile, and to
estimate $\mu$ at any distance. As for the central
magnitudes, we have not corrected $\mu$ for internal dust obscuration.

One important uncertainty in the determination of the central
magnitude and surface brightness, in particular the measurements at
0.2 arcsec, comes from the determination of the centroid of the
maximum central emission.  Several galaxies do not have a single
central maximum, rather they show several knots produced by
a patched central extinction or/and by the actual presence of several
clusters. In these cases, the magnitude and surface brightness listed
in Table 5 correspond to the measurements obtained placing the center
of the aperture at the centroid of the brightest knot. Obviously, for
these galaxies (NGC 660, NGC 841, NGC 3627, NGC 4192 and NGC 5005) the
central 0.2 arcsec magnitude can be significantly different, by up to
more than 0.1 mag, if the aperture is centered in another
knot. However, the integrated magnitudes at 0.5 and 1 arcsec radius
are not affected by the choice of the aperture center, even for those
galaxies with several central knots. The central measurements of the
galaxies that have a very obscured nucleus, such as NGC 2911, NGC
4150, NGC 3166, and NGC 7177, are affected by the choice of the galaxy
center.

As expected, the galaxies that have been observed at F547M and at
F555W show surface brightness profiles and magnitudes that are equal
in both filters. The magnitudes and surface brightness profiles in the
F606W filter can be slightly brighter than the same quantities
measured in the F547M or F555W filters. Five galaxies (NGC 3169, NGC
3627, NGC 5005, NGC 5055 and NGC 6951) have been observed in F547M (or
F555W) and F606W. Except for NGC 3627, the other four galaxies are
brighter by about 0.1 mag in F606W than in F547M or F555W. This
difference can be explained by the contribution of H$\alpha$ emission
in F606W that does not affect F547M or F555W. Nuclear spectra of these
galaxies show indeed H$\alpha$ in emission that can account for the
small difference in magnitude between the two filters.  Because this
difference in the F606W and F555W (or F547M) magnitude is small, for
statistical purposes we consider all three filters, F547M, F555W and
F606W as equivalent to a visual magnitude.

Figure 7 shows the distribution of the magnitudes of the galaxies
observed in any of the F547M, F555W, or F606W filters.  Table 7 shows
the mean magnitude and surface brightness for all the objects and
for those that fall into the categories of LINERs, TO, galaxies with central
young/intermediate age population, and galaxies with old stellar
population (following the $W_K$-based criterion explained in Section 1).  
Figure 7 and Table 7 indicate that LINERs and TO
have similar central magnitudes and surface brightness in the F547M,
F555W, or F606W bands.  However, galaxies with central
young/intermediate age populations are about 0.5 mag (or 0.5
mag/arcsec$^2$) brighter than galaxies with central old stellar
populations. Because we have not corrected these magnitudes by
internal dust obscuration, the true difference in brightness is
certainly {\em larger} than this, since these galaxies with
central young/intermediate age populations are also dustier 
(Section 5, Paper III). From the spectral analysis carried
out in Paper III, we find that the central regions of Young systems
have, on average, $A_V$ three times larger than Old systems (mean
$A_V$ of 0.6 and 0.2 mag, respectively). Therefore, correcting for
internal dust would only strengthen this result.

In the F814W band, the surface brightness at 0.2 arcsec in sources
with young/intermediate age populations is larger than in those with
old stellar populations. These differences disappear when the surface
brightness is measured at 1 arcsec. These results suggest that a
fraction of the LLAGNs with young/intermediate age populations may
have compact bright central sources.

\subsection{Frequency of nuclear compact sources}

HST observations have revealed that 50-70$\%$ of the spiral galaxies
contain nuclear stellar clusters (Carollo et al 1998, 2002; Boeker et
al 2002, 2004).  Compact nuclear sources have also been
detected by C\^ot\'e et al (2006) and Ferrarese et al (2006a) in
almost all early type galaxies in their survey, except in giant
ellipticals.  These compact sources are not associated with AGNs, and
most of them are extended. Their colors suggest that they are clusters
with old and intermediate age populations. Ferrarese et al (2006b)
argue that they are low-mass counterparts of the BHs detected in
bright elliptical galaxies.

Our goal here is  to determine the frequency of
nuclear compact sources in LLAGNs and identify possible differences between
their sub-types.  To determine the frequency of nucleated galaxies in
LLAGNs can be a key point to determine the nature of TOs and LINERs, and
the differences between both types of nuclear activity.   

A compact light source is identified sometimes for rising above the
inward extrapolated surface brightness cusp at small radii. For
example, if the underlying galaxy is modeled by a Nuker law, then
the nucleus is identified as an excess above the power law extrapolation
(e.g. Lauer et al. 1995, Ravindranath et al. 2001, Rest et al. 2001, 
Scarlata et al. 2004, Lauer et al. 2005, Hughes et al. 2005).
However, Graham et al. (2003) and Trujillo et al. (2004) have 
shown that a S\'ersic model (S\'ersic 1968) can provide a
better fit to the surface brightness profiles of early type galaxies,
since a Nuker law is quite sensitive to the radial extent of the
data. Recently, Ferrarese et al. (2006a) and C\^ot\'e et al. (2006)
have also shown that a S\'ersic, or a core-S\'ersic provide adequate
fits to the whole surface brightness profile of elliptical galaxies
in the Virgo cluster. This is actually a controversial topic, and
arguments favoring the Nuker (Lauer et al. 2006) or the S\'ersic
models (Ferrarese et al. 2006c) have been given extensively during the
last year.

In summary, an excesses above the Nuker, Sersic or core-Sersic
fits can be produced by a compact source. Note, however, that these
methods are able to identify a bright compact source at the galaxy
center, but do not determine the nature of the nuclear component, i.e.,
whether it is due to an AGN or a stellar cluster.  If the source is
extended,  this nuclear component is probably related with stellar
clusters. But, if the central source is spatially unresolved, and the
central light excess is well fitted by a PSF, we can not say too much
about the nature of the nucleated component. AGNs are expected to
be point sources, but certainly many nuclear stellar clusters at
the distance of these galaxies can be also unresolved (e.g. the
nuclear cluster in NGC 4303, Colina et al. 2003).
Due to these limitations, we propose to combine the morphology and surface
brightness profile analysis together with the study of the nuclear
stellar population properties of these objects (results already
obtained in Papers I, II and III) to try to distinguish between the AGN
or stellar cluster nature of the nuclear compact sources.

To identify whether a LLAGN has a compact source we have inspected the
surface brightness profiles of these galaxies. We use the task
'ellipse' in STSDAS to fit isophotes. If the galaxy
shows dust filaments, lanes or discs but the dust obscuration is not
severe, we used the following approach. First, for each galaxy, the
isophotes were constrained by fitting ellipses with a fixed center,
ellipticity and position angle, chosen as the best fit to the
apparent galaxy shape in the regions where dust is not present. A
model image was then built with the task "bmodel" in STSDAS using the
results from this first fit. This model image is subtracted from the
data, producing a dust frame that is used later to correct the
original image. A new fit is performed in the image corrected by
dust, but now the center, ellipticity and position angle of each
isophote were not constrained. The results for two galaxies with no
dust (e.g. NGC 3998) or with a small amount of obscuration (e.g. NGC
6384) are shown in Figure 8. Other examples (NGC 2787 and NGC 5337) are in
Figure 9. Here, the obscuration is larger but the dust is distributed
mainly in only one side of the galaxy, and we are able to perform the
fit to the undistorted part of the isophotes.

Unfortunately, many of the galaxies in this sample show non-elliptical
isophotes, and/or the dust obscuration is severe (e.g. NGC 4150, NGC
7177), so that we are not able to fit their light distribution. Hence,
the analysis of the surface brightness profile can not be the main
method used to find out the fraction of LLAGNs that have compact sources; a
different approach is needed to identify compact sources in these
galaxies. We have identified nuclear compact sources by visual
inspection of the HST images, as done by Carollo et al (1998). If a
color image is available, we also inspect the color frame to check if
a compact source with a color different from that of the underlying
background is detected (see, for example, the F547M/F814W image of NGC
5055 in Fig 3). Additionally, we have built a surface brightness
profile for all the galaxies of the sample by performing aperture
photometry. We have checked that the profiles built in this way are
quite similar to the profiles obtained using the task "ellipse" in
STSDAS for the galaxies that are well fitted by elliptical isophotes
(see the comparison for NGC 3998 and NGC 6384 in Figure 8). The
agreement between the profiles obtained with the two processes is
remarkable for most of the galaxies. Finally, for the objects that we
identify visually a central source, we explore if the surface
brightness also shows an inflexion point at small radii in the 
gradient profile. This gradient is obtained by calculating for each radius
the slope of the profile. Figure 8 shows the gradient profile obtained
for NGC 3998 and NGC 6384. Note, however, that by itself, an
inflection point in the light profile does not necessarily reflect the
presence of a nuclear component because it can also be produced by
varying dust obscuration.

Summarizing, we assume that the galaxy has a nuclear compact source if
the following conditions are met: we visually identified this source
in the image, and/or in the color image, in the surface brightness
profile (built fitting ellipses to the isophotes and/or by aperture
photometry) and by an inflexion point in the gradient profile at small
radii. Table 4 lists the LLAGNs that have nuclear compact sources.
Following Ferrarese et al (2006a), we call them nucleated
galaxies. This table also mentions the objects that have several
central sources. Usually these objects also show dust at the center,
and in these cases it is difficult to know whether they are several
real sources or a consequence of the dust distribution. 

Table 7 summarizes the results of this analysis, indicating the
frequency of LLAGNs that can be classified as nucleated, non-nucleated
and those with several compact sources and central dust.  The
frequency of nucleated LLAGNs (51\%) is similar to the fraction of
spiral galaxies with nuclear stellar clusters (Carollo et al. 1998,
2002; Boeker et al 2002, 2004). This result is not unexpected since
most (93\%) of the LLAGNs are spirals. We do not find any significant
difference in the frequency of nucleated strong- and weak-[OI]
LLAGNs. However, the fraction of galaxies classified as Young-TOs that
have compact nuclear sources (67\%) is much higher than those that are
non-nucleated (22\%). Old-TOs, on the other hand, are somewhat more
frequently non-nucleated (58\%). LINERs host mainly old populations,
and they are equally frequent in the nucleated and non-nucleated
categories. None of the only two LINERs classified as Young is
classified as nucleated, but both show knots (presumably stellar clusters) and dust at the
center. 

To estimate the nuclear magnitude, we have first estimated the maximum
radius of the nuclear component by inspecting the gradient profile and
the inflexion point. Then, the magnitude was estimated by measuring
the flux in a circular aperture of radius equal to the inflexion point
(r$_{in}$). This flux is subtracted from the underlying galactic light
contribution that is assumed to have equal surface brightness that the
annulus of inner (r$_{in}$+1 pixel) and outer (r$_{in}$+3 pixels)
radii. To estimate the uncertainty in the nuclear magnitude, we have
estimated this quantity by changing r$_{in}$ by $\pm1$ pixel, and by measuring the
background light contribution in annuli of radii (r$_{in}$,r$_{in}+2$)
and (r$_{in}+2$,r$_{in}+4$). Table 8 lists 
r$_{in}$ and the resulting measurements. This method to
estimate the nuclear magnitude is similar to the method used by Carollo
et al (1997), who fit the central source with a gaussian, and assume
that the underlying galactic continuum light is represented by the
asymptotic value of the gaussian wings.  In fact, our results for the
three galaxies (NGC 5377, NGC 5985 and NGC 6384) in common with the
Carollo et al (1998) sample agree with their values. This method also
provides similar (or lower) values to the nuclear magnitude
estimated fitting the surface brightness profile with a King model (King 1968) 
for the central source plus a Sersic law for the underlying galaxy.

Figure 10 shows the distribution of the nuclear magnitudes and surface
brightnesses of the nucleated galaxies observed in any of the V
(F547M, F555W, F606W) filters divided into emission-line (LINERs and
TO) and stellar-population groups. Despite the overlap, it is clear
that, as for the central magnitudes studied in Section 6.1, Young
systems are brighter. The average difference is about 2 mag in the
nuclear surface brightness. Again, this should be considered a lower
limit, since we have not corrected for internal extinction, and Young
systems are dustier than Old ones. Furthermore, most of the
galaxies known to have some contribution from the AGN to the nuclear
component (NGC\,315, NGC\,3998, NGC\,4203 and NGC\,4261) are LINERs.
If it were possible to take the AGN contribution into account, one would
find a larger difference in the magnitudes of the nuclear stellar
components.

In order to further investigate the nature of the nuclear components in this
sample we compare our results with those from Chiaberge, Capetti \&
Macchetto (2005). These authors compared, in Figure 4 of their paper,
the distribution of nuclear radio and optical luminosities of LINER's
with those of FRI radio galaxies and Seyfert galaxies. They find that
a small number of LINER's follow in the relation observed for Radio
galaxies, indicating that in their case the nuclear component is due
to synchrotron radiation at the base of the jet (Chiaberge et al. 1999).
Two of the galaxies in our sample (NGC\,315 and NGC\,4261) fall in this
category. They also found that the remaining galaxies fall closer to the
relation defined by Seyfert galaxies. 

In Figure~11 we compare our galaxies with the Seyfert's and LINER's from
Chiaberge et al. (2005)\footnote{Notice that Chiaberge et al. (2005) used 
the classification from Ho et al. 1997, which is slightly different from the
one used here. As a result some of their galaxies (e.g. NGC404, NGC 3368, 
NGC 4314 and NGC 4736) should be considered TO's.}. 
For simplicity we excluded the Radio Galaxies.
Radio fluxes obtained from Nagar, Falcke, \& Wilson (2005), and are
presented in Table~7. These measurements were converted from 15~GHz
to 5~GHz assuming a flat spectral index. We also converted the optical
luminosities of our compact nuclear sources to $\lambda=$7000\AA, the
same wavelength used by Chiaberge et al. (2005), assuming a spectral
index $\nu^{-1}$. We can see in this Figure that in the case of TO's,
either Old or Young ones, 15 out of the 17 sources were not detected
in radio. On the other hand only 1 of the 9 LINER's was not detected.

The distribution of our sources in Figure~11 shows an interesting result.
We find that the Old-LINER's follow the relation defined by Seyfert galaxies,
extending it to lower luminosities and suggesting that the nuclear
component in these sources could be due to AGN emission. Further evidence in 
favor of this interpretation comes from the detection of broad H$\alpha$
emission in most of these sources (see Table 1) (Ho et al. 2000). In the case of TO's,
in particular the Young ones, their distribution is displaced relative
to the relation defined by Seyferts and LINER's. These galaxies have
higher optical luminosities than what one would expect based on their
radio luminosities. This result suggests that the source of the optical
emission is different from that of Old-LINER's, and most likely have a
stellar origin, consistent with the spectroscopic results. 

While a detailed interpretation of these findings is beyond the scope
of this paper, we close this section with a speculation on a possible
evolutionary scenario. Assuming TO nuclei to be stellar clusters,
their detectability above the underlying background depends on its
brightness, which, for a given mass, decreases with age\footnote{However, 
the level of detectability also depends on the surface brightness
profile of the underlying galaxy. So, a compact source is more easily
detected in a galaxy with a flat surface brightness profile that in
one with a steep power law cusp.}. Given the
difficulties in identifying low contrast nuclei, it is likely that a
cluster which is Young and detected now would fade beyond
detection when it gets Old. Furthermore, as it ages, its color becomes
more similar to that of the underlying old stellar population, 
and thus more difficult to be identified with our color criterion. 
Dust present at Young ages
should tend to dissipate or settle onto a more organized structure.
Between the age ranges typical of our Young and Old
stellar-population categories (0.1--1 Gyr and 10 Gyr, respectively) a
coeval stellar population fades by $\sim 2$--4 mag. The observed
change in brightness will be smaller since extinction decreases with
time. Using the typical $0.4$ mag difference in $A_V$ between Young
and Old LLAGN estimated in Paper III, give brightness changes between
1.6 and 3.6 mag, depending on whether we use 1 or 0.1 Gyr to characterize
the Young class. Needless to say, these are very rough numbers, but
they do compare favorably with the average difference of $\sim 2$ mag
in nuclear surface brightness between Young and Old systems.

A more detailed analysis should quantify the
detectability threshold and take into account the ranges of masses and
ages for such clusters, as well as the possibility that they have not
been formed instantaneously (van der Marel et al. 2007). At
present, however, this qualitative evolutionary scenario seems
consistent with most of our findings.

\section{Summary and Conclusions}

LLAGNs, that include LINERs and TOs, are the most common type of
AGN. What powers them is still at the forefront of AGN research.  To
unveil the nature of the central source we are constructing a
panchromatic atlas of the inner regions of these galaxies, which will
be used to determine their nuclear stellar population.  To this end we
have already carried out a near-UV snapshot survey of nearby LLAGNs
with the ACS on board HST, that is complemented with optical and near-IR
images available in the HST archive.

In this paper, the first of a series, we present observations of 57
LLAGNs imaged with the WFPC2 through any of the V (F555W, F547M,
F606W) and/or I (F791W, F814W) bands. These objects comprise 36$\%$ of
the original HFS97 LLAGNs sample, and correspond to those for which
there are WFPC2 images available in the HST archive and whose
circumnuclear stellar population we have already studied
spectroscopically (Papers I--III). The subset of objects studied here
follows the same distance and morphological type distribution of the
complete HFS97 LLAGN sample. Classifying the objects in strong-[OI]
and weak-[OI] ([OI]/H$\alpha$)$\leq$ 0.25), this subset includes
34$\%$ and 36$\%$ of the strong- and weak-[OI], respectively, of the
whole HFS97 LLAGN sample. Following our results obtained from the
analysis of the circumnuclear stellar population, this sub-sample
contains 17 Young-TOs, 20 Old-TOs, 18 Old-LINERs and 2
Young-LINERs. Young-TOs or Young-LINERs are LLAGNs in which
intermediate age stars contribute significantly to the nuclear
blue-optical continuum. The dearth of Young-LINERs in the sample can
be understood from the result obtained in our spectroscopic studies, which
show that the overwhelming majority of LINERs harbor an old stellar
population. 

With these data we have built an atlas that includes the structural map
for all the images, and color maps for those galaxies for which
images in two filters are available. We have
identified those galaxies that have nuclear compact sources, and we
have studied the circumnuclear environment of LLAGNs.  We have found
circumnuclear dust in 88$\%$ of the LLAGNs, but this fraction is
somewhat larger (95$\%$) in weak-[OI] LLAGNs. The dust morphology is
quite complex, from nuclear spiral lanes, to chaotic filaments and to
nuclear disk-like structures. Chaotic filaments are as frequent as the
well organized dust spirals; but disks are mainly seen in strong-[OI]
LLAGNs. The dust concentration (simply graded 
by its location relative to a radius of 100-200 pc, 
 is similar in weak- and in strong-[OI] because
the fraction of LLAGNs with dust located only in the inner part is 
larger in Old-LLAGNs than in Young-LLAGNs. These results suggest an evolutionary
dust sequence from Young-TOs to Old-LLAGNs. 

We have found that LINERs and TOs have both similar central magnitude and
surface brightness, but LLAGNs with young and intermediate age
populations are brighter than Old-TOs and LINERs. We have not found
any correlation between the presence of nuclear compact sources and
the emission line spectral type, ie., LINERs are as frequently
nucleated as TOs.  However, the centers of Young-TOs are brighter than
the centers of Old-TOs and LINERs. The difference in magnitude and
surface brightness can be even larger if we account for internal
extinction, since Young-TOs are dustier. This result indicates that 
Young-TOs are separated from other type of LLAGNs also in terms 
of their central brightness, in addition of the properties and spatial 
distribution of the stellar population. 

These data have been very useful to study the circumnuclear environment of LLAGNs, and to
identify which of these galaxies have a nuclear compact source. 
The fact that compact sources are as frequent in LINERs as in TOs,
confirms again that LLAGNs are a mixed bag of objects. These results also
suggest that the central morphology alone is not sufficient to elucidate 
the origin of their central source, and it cannot be used to ascertain whether
 LLAGNs are powered by AGNs or stellar clusters.  
These data will be complemented with near-UV (ACS) and near-IR
(NICMOS) images to provide a panchromatic atlas of the inner regions
of LLAGNs and to further investigate the origin of the nuclear sources
and their relation with stellar clusters.

{\bf Acknowledgements} We thanks the referee for her/his suggestions that helped to improve the paper. RGD and EP acknowledge support from the Spanish
Ministerio de Educaci\'on y Ciencia through the grant
AYA2004-02703. The data used in this work come from observations made
with NASA/ESA Hubble Space Telescope, obtained from the STScI data
archive. Basic research in radio astronomy at the NRL is supported by
6.1 base funding. We also thank support from a joint CNPq-CSIC
bilateral collaboration grant.



\begin{deluxetable}{lrrrrrrrrrr}
\tabletypesize{\tiny}
\tablewidth{0pc}
\tablecaption{Sample Properties}
\tablehead{
  \colhead{Galaxy}   &
  \colhead{Spectral} &
  \colhead{Hubble}   &
  \colhead{Type}   &
  \colhead{vel.}        &
  \colhead{dist.}    &
  \colhead{pc/$^{\prime\prime}$} &
  \colhead{[OI]/H$\alpha$} & 
  \colhead{W$_K$}  &
  \colhead{Dn4000}  &
   \colhead{SP Class}
  \\
  \colhead{Name}     &
  \colhead{Class}    &
  \colhead{Type}     &
  \colhead{Type}     &
  \colhead{km$\,$s$^{-1}$} &
  \colhead{Mpc}      &
  \colhead{pc}       &
  \colhead{} &
  \colhead{\AA} &
 \colhead{} &
  \colhead{}         
  }
\startdata
NGC266   &   L1.9   &   SB(rs)ab      &   2.0  &   4662  &   62.4  &   302  &   0.28  &   18.82  &   2.03  &   OL \nl
NGC315   &   L1.9   &   E+:           &   -4.0 &   4935  &   65.8  &   319  &   0.59  &   17.01  &   2.00  &   OL \nl
NGC404   &   L2     &   SA(s)0-:      &   -3.0 &   -46 &   2.4  &   11  &   0.17  &   9.82  &   1.50  &   YT \nl
NGC428   &   L2/T2: &   SAB(s)m       &   9.0  &   1160  &   14.9  &   72  &   0.19  &   14.83  &   1.65  &   YT  \nl
NGC660   &   T2/H:  &   SB(s)apec     &   1.0  &   852  &   11.8  &   57  &   0.047  &   14.53  &  1.72  &   YT \nl
NGC841   &   L1.9:  &   (R')SAB(s)ab  &   2.1  &   4539  &   59.5  &   288  &   0.58  &   14.93  &   1.73  &   YL \nl
NGC1161  &   T1.9:  &   SA0           &   -2.0 &   1940  &   25.9  &   125  &   0.14  &   18.98  &   2.16  &   OT \nl
NGC2685  &   S2/T2: &   (R)SB0+pec    &   -1.0 &   820  &   16.2  &   78  &   0.13  &   18.72  &   2.01  &   OT \nl
NGC2787  &   L1.9   &   SB(r)0+       &   -1.0 &   691  &   13.0  &   63  &   0.55  &   16.82  &   2.08  &   OL \nl
NGC2911  &   L2     &   SA(s)0:pec    &   -2.0 &   3183  &   42.2  &   204  &   0.31  &   17.72  &   2.05  &   OL \nl
NGC3166  &   L2     &   SAB(rs)0/a    &   0.0  &   1344  &   22  &   106  &   0.27  &   15.88  &   1.80  &   OL \nl
NGC3169  &   L2     &   SA(s)apec     &   1.0  &   1234  &   19.7  &   95  &   0.28  &   17.31  &   1.99  &   OL \nl
NGC3226  &   L1.9   &   E2:pec        &   -5.0 &   1321  &   23.4  &   113  &   0.59  &   18.51  &   2.13  &   OL \nl
NGC3245  &   T2:    &   SA(r)0?       &   -2.0 &   1348  &   22.2  &   107  &   0.086  &   15.21  &   1.79  &   OT \nl
NGC3368  &   L2     &   SAB(rs)ab     &   2.0  &   897  &   8.1  &   39  &   0.18  &   10.12  &   1.48  &   YT \nl
NGC3489  &   T2     &   SAB(rs)0+     &   -3.0 &   701  &   6.4  &   31  &   0.11  &   12.98  &   1.58  &   YT \nl
NGC3507  &   L2     &   SB(s)b        &   3.0  &   978  &   19.8  &   96  &   0.18  &   5.44  &   1.26  &   YT \nl
NGC3627  &   T2/S2  &   SAB(s)b       &   3.0  &   727  &   6.6  &   31  &   0.13  &   11.61  &   1.55  &   YT \nl
NGC3675  &   T2     &   SA(s)b        &   3.0  &   766  &   12.8  &   9  &   0.12  &   18.45  &   2.17  &   OT \nl
NGC3705  &   T2     &   SAB(r)ab      &   2.0  &   1018  &   17.0  &   82  &   0.079  &   14.8  &   1.73  &   YT \nl
NGC3992  &   T2:    &   SB(rs)bc      &   4.0  &   1048  &   17.0  &   82  &   0.13  &   18.69  &   2.36  &   OT \nl
NGC3998  &   L1.9   &   SA0           &   -2.0 &   1049  &   21.6  &   105  &   0.53  &   3.89  &   1.25  &   OL \nl
NGC4143  &   L1.9   &   SAB(s)0       &   -2.0 &   783  &   17.0  &   82  &   0.71  &   14.85  &   1.72  &   OL \nl
NGC4150  &   T2     &   SA(r)0?       &   -2.0 &   43  &   9.7  &   47  &   0.13  &   12.57  &   1.55  &   YT \nl
NGC4192  &   T2     &   SAB(s)ab      &   2.0  &   -142 &   16.8  &   81  &   0.14  &   15.95  &   1.93  &   OT \nl
NGC4203  &   L1.9   &   SAB0-         &   -3.0 &   1085  &   9.7  &   47  &   1.22  &   10.01  &   1.42  &   OL \nl
NGC4261  &   L2     &   E2+           &   -5.0 &   2210  &   35.1  &   170  &   0.49  &   18.45  &   1.92  &   OL \nl
NGC4314  &   L2     &   SB(rs)a       &   1.0  &   962  &   9.7  &   47  &   0.18  &   15.09  &   2.02  &   OT \nl
NGC4321  &   T2     &   SAB(s)bc      &   4.0  &   1234  &   16.8  &   81  &   0.11  &   7.49  &   1.32  &   YT \nl
NGC4429  &   T2     &   SA(r)0+       &   -1.0 &   1137  &   16.8  &   81  &   0.097  &   15.79  &   1.95  &   OT \nl
NGC4435  &   T2     &   SB(s)0        &   -2.0 &   781  &   16.8  &   81  &   0.13  &   15.06  &   1.96  &   OT \nl
NGC4438  &   L1.9   &   SA(s)0/a:     &   0.0  &   64  &   16.8  &   81  &   0.27  &   17.85  &   2.00  &   OL \nl
NGC4450  &   L1.9   &   SA(s)ab       &   2.0  &   1956  &   16.8  &   81  &   0.67  &   11.74  &   1.52  &   OL \nl
NGC4459  &   T2     &   SA(r)0+       &   -1.0 &   1202  &   16.8  &   81  &   0.13  &   14.83  &   2.21  &   YT \nl
NGC4569  &   T2     &   SAB(rs)ab     &   0.0  &   -235 &   16.8  &   81  &   0.062  &   4.99  &   1.21  &   YT \nl
NGC4596  &   L2::   &   SB(r)0+       &   -1.0 &   1874  &   16.8  &   81  &   0.27  &   15.9  &   2.05  &   OL \nl
NGC4736  &   L2     &   (R)SA(r)ab    &   2.0  &   307  &   4.3  &   20  &   0.24  &   12.93  &   1.67  &   YT \nl
NGC4826  &   T2     &   (R)SA(rs)ab   &   2.0  &   411  &   4.1  &   19  &   0.073  &   14.43  &   1.76  &   YT \nl
NGC5005  &   L1.9   &   SAB(rs)bc     &   4.0  &   948  &   21.3  &   103  &   0.65  &   14.6  &   1.74  &   YL \nl
NGC5055  &   T2     &   SA(rs)bc      &   4.0  &   504  &   7.2  &   34  &   0.17  &   14.07  &   1.75  &   YT \nl
NGC5377  &   L2     &   (R)SB(s)a     &   1.0  &   1792  &   31.0  &   150  &   0.25  &   8.66  &   1.45  &   YT \nl
NGC5678  &   T2     &   SAB(rs)b      &   3.0  &   1924  &   35.6  &   172  &   0.079  &   8.7  &   1.48  &   YT \nl
NGC5879  &   T2/L2  &   SA(rs)bc?     &   4.0  &   772  &   16.8  &   81  &   0.16  &   16.3  &   1.72  &   OT \nl
NGC5970  &   L2/T2: &   SB(r)c        &   5.0  &   1965  &   31.6  &   153  &   0.18  &   18.39  &   1.86  &   OT \nl
NGC5982  &   L2::   &   E3            &   -5.0 &   2904  &   38.7  &   187  &   0.49  &   18.11  &   2.05  &   OL \nl
NGC5985  &   L2     &   SAB(r)b       &   3.0  &   2518  &   39.2  &   190  &   0.30  &   18.9  &   1.97  &   OL \nl
NGC6340  &   L2     &   SA(s)0/a      &   0.0  &   1207  &   22.0  &   106  &   0.43  &   20.04  &   2.26  &   OL \nl
NGC6384  &   T2     &   SAB(r)bc      &   4.0  &   1667  &   26.6  &   128  &   0.15  &   18.6  &   1.92  &   OT \nl
NGC6500  &   L2     &   SAab:         &   1.7  &   2999  &   39.7  &   192  &   0.23  &   15.81  &   1.74  &   OT \nl
NGC6503  &   T2/S2: &   SA(s)cd       &   6.0  &   42  &   6.1  &   29  &   0.08  &   9.49  &   1.36  &   YT \nl
NGC6703  &   L2::   &   SA0-          &   -2.5 &   2364  &   35.9  &   174  &   0.36  &   18.53  &   2.11  &   OL \nl
NGC6951  &   S2     &   SAB(rs)bc     &   4.0  &   1424  &   24.1  &   116  &   0.23  &   16.4  &   1.70  &   OT \nl
NGC7177  &   T2     &   SAB(r)b       &   3.0  &   1147  &   18.2  &   88  &   0.14  &   16.64  &   1.80  &   OT \nl
NGC7217  &   L2     &   (R)SA(r)ab    &   2.0  &   945  &   16.0  &   77  &   0.25  &   19.2  &   2.07  &   OT \nl
NGC7331  &   T2     &   SA(s)b        &   3.0  &   821  &   14.3  &   69  &   0.097  &   18.03  &   1.99  &   OT \nl
NGC7626  &   L2::   &   E:pec         &   -5.0 &   3422  &   45.6  &   221  &   0.22  &   18.08  &   2.12  &   OT \nl
NGC7742  &   T2/L2  &   SA(r)b        &   3.0  &   1653  &   22.2  &   107  &   0.13  &   17.07  &   1.90  &   OT \nl
\hline
\enddata
\label{tab:sample_properties}
\tablecomments{Col.\ (1): Galaxy name; Col.\ (2): Spectral class. Col.\
(3): Hubble type. Col.\ (4): Numerical Hubble type. Col.\ (5): Radial
velocity.  Col.\ (6): Distance. Col.\ (7): Angular scale.  Col.\ (8):
[OI]/H$\alpha$ flux ratio. All these quantities were extracted from HFS97.
Col.\ (9): Equivalent width of CaII K line. Col.\ (10): The 4000 \AA\
break. Col.\ (11): Stellar population class. O: old, Y: young population.
These three columns are from Cid Fernandes et al. (2004) and Gonz\'alez
Delgado et al. (2004).}
\tablenotetext{1,2}{Grouped among TOs in this paper.}  
\tablenotetext{3}{Grouped with LINERs in this paper (see P\'erez \etal 2000).}
\end{deluxetable}

\begin{deluxetable}{lrrrrrrrrrr}
\tabletypesize{\tiny}
\tablewidth{0pc}
\tablecaption{Observations}
\tablehead{
  \colhead{Galaxy}   &
   \colhead{Filter}   &
  \colhead{Texp} &
  \colhead{scale}   &
  \colhead{orient}        &
  \colhead{ID}    &
   \colhead{Filter}   &
 \colhead{Texp} &
  \colhead{scale}   &
  \colhead{orient}        &
  \colhead{ID}  
  \\
  \colhead{Name}     &
   \colhead{}     &
  \colhead{s}    &
  \colhead{arcsec/pix}     &
  \colhead{deg} &
  \colhead{}      &
    \colhead{}      &
   \colhead{s}    &
  \colhead{arcsec/pix}     &
  \colhead{deg} &
  \colhead{}   
  }
\startdata
NGC 266		&	F547M	&	360	&	0.0456	&	-56.5		&	6837	&		&				&			&           &        \nl
NGC 315		&	F547M	&	360	&	0.0456	&	-55.7		&	6837	&	F814W	&	460.0	&	0.0456	&	97.10	&	6673 \nl
        			&	F555W	&	460	&	0.0456	&	97.1		&	6673	&		&		&		&		&	\nl
NGC 404		&	F547M	&	350	&	0.0456	&	-156.0	&	6871	&	F814W	&	320.0	&	0.0456	&	-92.90	&	5999 \nl
    			&	F555W	&	160	&	0.0456	&	-92.9		&	5999	&		&		&		&		&	\nl
NGC 428		&	F606W	&	460	&	0.100	&	-157.7	&	9042	&	F814W	&	640.0	&	0.0456	&	112.0	&	8599 \nl
NGC 660		&	F606W	&	160	&	0.100	&	-155.1	&	5446	&	F814W	&	460.0	&	0.100	&	-154.9	&	9042 \nl
NGC 841		&	F547M	&	400	&	0.0456	&	117.0	&	6837	&		&		&		&		&	\nl
NGC 1161 	&	F547M	&	360	&	0.0456	&	-113.0	&	6837	&		&		&		&		&	\nl
NGC 2685 	&	F555W	&	1000	&	0.0456	&	-135.0	&	6633	&	F814W	&	730.0	&	0.0456	&	-135.0	&	6633 \nl
NGC 2787 	&	F547M	&	360	&	0.0456	&	26.8		&	6837	&	F814W	&	730.0	&	0.0456	&	-125.0	&	6633 \nl
    	 		&	F555W	&	1000	&	0.0456	&	-125.0	&	6633	&		&		&		&		&	\nl
NGC 2911 	&	F547M	&	460	&	0.100	&	-115.6	&	5924	&		&		&		&		&	\nl
NGC 3166 	&	F547M	&	300	&	0.0456	&	-29.0		&	5419	&		&		&		&		&	\nl
NGC 3169 	&	F547M	&	260	&	0.0456	&	-25.9		&	5419	&	F814W	&	460.0	&	0.100	&	-115.8	&	9042 \nl
NGC 3226 	&	F547M	&	460	&	0.0456	&	173.0	&	6837	&		&		&		&		&	\nl
NGC 3245 	&	F547M	&	360	&	0.0456	&	178.0	&	6837	&		&		&		&		&	\nl
NGC 3368 	&	F606W	&	320	&	0.100	&	-110.8	&	9042	&	F814W	&	320.0	&	0.100	&	-110.8	&	9042 \nl
NGC 3489 	&	F555W	&	100	&	0.0456	&	-31.7		&	5999	&	F814W	&	200.0	&	0.0456	&	-31.70	&	5999 \nl
NGC 3507 	&	F606W	&	160	&	0.0456	&	151.0	&	5446	&		&		&		&		&	\nl
NGC 3627 	&	F606W	&	560	&	0.0456	&	-20.1		&	8597	&		&		&		&		&	\nl
NGC 3675 	&	F606W	&	160	&	0.100	&	-119.4	&	5446	&		&		&		&		&	\nl
NGC 3705 	&	F606W	&	160	&	0.100	&	64.3		&	5446	&	F814W	&	460.0	&	0.100	&	-107.6	&	9042 \nl
NGC 3992 	&	F547M	&	230	&	0.0456	&	149.0	&	5419	&		&		&		&		&	\nl
NGC 3998	&	F547M	&	240	&	0.0456	&	-128.0	&	5924	&	F791W	&	100.0	&	0.0456	&	-128.0	&	5924 \nl
NGC 4143	&	F606W	&	560	&	0.0456	&	-77.2		&	8597	&		&		&		&		&	\nl
NGC 4150	&	F555W	&	160	&	0.0456	&	-40.4		&	5999	&	F814W	&	320.0	&	0.0456	&	-40.40	&	5999 \nl
NGC 4192	&	F547M	&	720	&	0.0456	&	149.0	&	6436	&	F814W	&	660.0	&	0.0456	&	167.0	&	5375 \nl
  		  	&	F555W	&	660	&	0.0456	&	167.0	&	5375	&	F791W	&	720.0	&	0.0456	&	149.0	&	6436 \nl
NGC 4203	&	F547M	&	300	&	0.0456	&	11.9		&	5419	&	F814W	&	320.0	&	0.0456	&	-12.20	&	5999 \nl
   		 	&	F555W	&	160	&	0.0456	&	-12.2		&	5999	&		&		&		&		&	\nl
NGC 4261	&	F547M	&	800	&	0.0456	&	-20.3		&	5124	&	F791W	&	800.0	&	0.0456	&	-20.30	&	5124 \nl
NGC 4314	&	F606W	&	560	&	0.0456	&	-10.7		&	8597	&	F814W	&	600.0	&	0.0456	&	-5.310	&	6265 \nl
NGC 4321	&	F555W	&	1668	&	0.0456	&	-26.1		&	5195	&		&		&		&		& \nl	
NGC 4429	&	F606W	&	160	&	0.0456	&	151.0	&	5446	&		&		&		&		&	\nl
NGC 4435	&				&		&		&		&		&	F814W	&	520.0	&	0.0456	&	-93.40	&	6791 \nl
NGC 4438	&				&		&		&		&		&	F814W	&	1050	&	0.100	&	-169.7	&	6791 \nl
NGC 4450	&	F555W	&	460	&	0.0456	&	175.0	&	5375	&	F814W	&	460.0	&	0.0456	&	175.0	&	5375 \nl
NGC 4459	&	F555W	&	160	&	0.0456	&	-111.0	&	5999	&	F814W	&	320.0	&	0.0456	&	-111.0	&	5999 \nl
NGC 4569	&	F547M	&	126	&	0.0456	&	-164.0	&	6436	&	F791W	&	126.00	&	0.0456	&	-164.0	&	6436 \nl
NGC 4596	&	F606W	&	160	&	0.0456	&	153.0	&	5446	&		&		&		&		&	\nl
NGC 4736	&	F555W	&	296	&	0.0456	&	23.3		&	5741	&		&		&		&		&	\nl
NGC 4826	&	F547M	&	1600	&	0.0456	&	-140.0	&	8591	&		&		&		&		&	\nl
NGC 5005	&	F547M	&	230	&	0.0456	&	161.0	&	5419	&	F791W	&	120.0	&	0.0456	&	-168.0	&	6436 \nl
   			&	F606W	&	560	&	0.0456	&	167.0	&	8597	&	F791W	&	600.0	&	0.0456	&	-168.0	&	6436 \nl
NGC 5055	&	F606W	&	160	&	0.0456	&	161.0	&	5446	&	F814W	&	460.0	&	0.100	&	-78.09	&	9042 \nl
			&	F547M	&	1050	&	0.0456	&	-0.04		&	8591	&		&		&		&		&	\nl
NGC 5377	&	F606W	&	600	&	0.0456	&	-156.0	&	6359	&		&		&		&		&	\nl
NGC 5678	&	F606W	&	600	&	0.0456	&	-103.0	&	6359	&		&		&		&		&	\nl
NGC 5879	&	F606W	&	600	&	0.0456	&	157.0	&	6359	&	F814W	&	170.0	&	0.0456	&	77.20	&	7450 \nl
NGC 5970	&	F606W	&	560	&	0.0456	&	179.0	&	8597	&		&		&		&		&	\nl
NGC 5982	&	F555W	&	1000	&	0.0456	&	-134.0	&	5454	&	F814W	&	460.0	&	0.0456	&	-134.0	&	5454 \nl
NGC 5985	&	F606W	&	600	&	0.0456	&	-68.4		&	6359	&		&		&		&		&	\nl
NGC 6340	&	F606W	&	600	&	0.0456	&	-129.0	&	6359	&		&		&		&		&	\nl
NGC 6384	&	F606W	&	600	&	0.0456	&	-83.3		&	6359	&		&		&		&		&	\nl
NGC 6500	&	F547M	&	350	&	0.0456	&	-12.2		&	5419	&		&		&		&		&	\nl
NGC 6503	&	F606W	&	160	&	0.0456	&	-178.0	&	5446	&		&		&		&		&	\nl
NGC 6703	&	F555W	&	160	&	0.0456	&	68.1		&	5999	&	F814W	&	320.0	&	0.0456	&	68.10	&	5999 \nl
NGC 6951	&	F606W	&	560	&	0.0456	&	114.0	&	8597	&	F814W	&	700.0	&	0.100	&	43.16	&	8602 \nl
			&	F547M	&	300	&	0.0456	&	-123.0	&	5419	&		&		&		&		&	\nl
NGC 7177	&	F606W	&	560	&	0.0456	&	136.0	&	8597	&	F814W	&	460.0	&	0.100	&	-176.7	&	9042 \nl
NGC 7217	&	F547M	&	300	&	0.0456	&	-73.0		&	5419	&	F814W	&	460.0	&	0.100	&	-178.0	&	9042 \nl
			&	F547M	&	300	&	0.0456	&	-73.0		&	5419	&		&		&		&		&	\nl
NGC 7331	&			&			&			&		&		&	F814W	&	170.0	&	0.0456	&	-115.0	&	7450 \nl
NGC 7626	&	F555W	&	1000	&	0.0456	&	-65.5		&	5454	&	F814W	&	460.0	&	0.0456	&	-65.50	&	5454 \nl
NGC 7742	&	F555W	&	480	&	0.100	&	-163.1	&	6276	&	F814W	&	680.0	&	0.100	&	-163.1	&	6276 \nl

\hline
\enddata
\label{tab:observations}
\end{deluxetable}


\begin{deluxetable}{lrrrl}
\tabletypesize{\tiny}
\tablewidth{0pc}
\tablecaption{Dust and nuclear morphology}
\tablehead{
  \colhead{Galaxy}  &
  \colhead{dust}  &
   \colhead{dust} &
  \colhead{nucleated }   &
  \colhead{Comments}    
  \\
  \colhead{Name}     &
  \colhead{morph}    &
  \colhead{distr}     &
  \colhead{} &
  \colhead{}      
   }
\startdata
NGC 266  &       LW   &  inout   &   Y      &   Spiral dust lanes, a small nuclear dust disk?                \nl 
NGC 315  &       D    &  in      &   Y          &   Nuclear dust disk and dust chaotic filaments, a nuclear compact source                  \nl
NGC 404  &       CS   &  in      &   Y         & Nuclear stellar cluster, spiral and chaotic filaments                      \nl
NGC 428  &       N    &  --      &   Y            &    Two nuclear stellar clusters, the brightest one at the center of the galaxy                   \nl
NGC 660  &       LW   &  inout   &  no      & Spiral dust lane structure,  the position of the center is not clear                       \nl
NGC 841  &       GD   &  inout   &  C+D   &    A spiral dust lane down to the center, several central sources?             \nl
NGC 1161  &      TW   &  inout   &  no?   & Well organized spiral dust lanes, an extended nucleus  ?                   \nl
NGC  2685  &      C   &  inout   &  no?    &   Chaotic dust morphology at the NE, a strong nuclear stellar disk                   \nl
NGC  2787  &      TW  &  inout   &  Y      &   Dust ring perpendicular to the galaxy major axis, a nuclear central source               \nl
NGC  2911  &      C   &  inout   &  no      &    Highly obscured center with chaotic filaments              \nl
NGC  3166  &      CS  &  inout   &  no     &   Very perturbed central morpholoy, chaotic dust lanes down the center             \nl
NGC  3169  &      CS  &  inout   &  no     &    Chaotic spiral dust lanes down to the center                 \nl
NGC  3226  &      D   &  in      &  Y           &    Dust disk and a nuclear source                  \nl
NGC  3245  &      D   &  in      &  Y           &     Dust disk and a nuclear source                  \nl
NGC  3368  &      CS  &  inout   &   Y?    &  Perturbed center, with chaotic spiral dust lanes, a compact  nuclear source               \nl
NGC  3489  &      CS  &  in      &   Y        &  Chaotic spiral dust lanes down the center, a stellar disk and/or a nuclear compact source                     \nl
NGC  3507  &      C   &  in      &   Y         & Some chaotic dust filaments, a nuclear stellar cluster                    \nl
NGC  3627  &      C   &  inout   &  C+D  &      Chaotic dust lanes, several sources at the center        \nl
NGC  3675  &      TW  &  inout   &  no    &  Spiral dust lanes out to large distance                     \nl
NGC  3705  &      GD  &  inout   &  no?  &  Spiral dust lanes toward the center, maybe with a stellar disk or an extended nuclear source                    \nl
NGC  3992  &      C   &  in      &  no         & Only a few dust filaments around, light profile follows a r$^{1/4}$                 \nl
NGC  3998  &      N   &  --      &  Y           &  Almost no dust around, a bright nuclear compact source                    \nl 
NGC  4143  &      LW  &  inout   &  no?  &  Weak coherent  spiral dust lanes, maybe a weak nuclear compact source                   \nl
NGC  4150  &      C   &  in      &  no         &    Highly perturbed center, chaotic dust lanes across the nucleus             \nl
NGC  4192  &      CS  &  inout   &  C+D &    Highly inclined galaxy, chaotic dust and one or several compact sources not  at the center          \nl
NGC  4203  &      TW  &  in      &  Y        & Some weak spiral dust lane around a bright nuclear compact source                     \nl
NGC  4261  &      D   &  in      &  Y          &    Nuclear dust disk, and a weak nuclear compact source                \nl
NGC  4314  &      LW  &  inout   &  no     &  Star formation ring, spiral dust lanes associated to the ring                     \nl
NGC  4321  &      LW  &  inout   &  Y      &  Star formation in arms spiraling toward the nucleus with dust lanes associated, a compact nuclear cluster                      \nl
NGC  4429  &      LW  &  inout   &  Y      &   Well structured spiral ring, a compact source and/or a small stellar disk                    \nl
NGC  4435  &      D   &  in      &  no        &    High inclined dust disk             \nl
NGC  4438  &      C   &  inout   &  no     &   Very perturbed central morphology, chaotic dust lanes across the center                 \nl
NGC  4450  &      C   &  inout   &  Y      &  Chaotic dust lanes, a bright nuclear compact source                    \nl
NGC  4459  &      TW  &  inout   &  Y?  &  Dust spiral ring, a nuclear source                    \nl
NGC  4569  &      C   &  inout   &  Y      &   Chaotic dust lanes, a nuclear stellar cluster                    \nl
NGC  4596  &      N   &  out     &  Y      &     Very faint dust spiral structure, a nuclear compact source                  \nl
NGC  4736  &      TW  &  inout   &  Y      &  Spiral dust lanes down the nucleus, a compact nuclear stellar cluster                    \nl
NGC  4826  &      TW  &  in      &  no     &    Inner spiral arms with star formation, dust associated to the arms                 \nl
NGC  5005  &      CS  &  inout   & D+C     &    Chaotic nuclear spirals, center is obscured but several central knots are visible           \nl
NGC  5055  &      C   &  in      &  Y      &    Dust is mainly in the S part, probably an extended nuclear source              \nl
NGC  5377  &      TW  &  inout   &  Y?     & Spiral dust lanes toward the nucleus, an extended nuclear source                 \nl
NGC  5678  &      CS  &  inout   &  Y      &   Chaotic dust lanes toward the center, a nuclear source              \nl
NGC  5879  &      LW  &  inout   &  Y      &  Weak nuclear source                    \nl
NGC  5970  &      LW  &  inout   &  Y      &   Spiral dust lanes, a faint compact nuclear source                  \nl
NGC  5982  &      N   &  --      &  no     &     No dust, a core light distribution                   \nl
NGC  5985  &      C   &  inout   &  Y?     &  A faint nuclear stellar cluster                   \nl
NGC  6340  &      N   &  --      &  no     &  Light profile follows a r$^{1/4}$                \nl
NGC  6384  &      CS  &  inout   &  Y      &  Dust lane toward the nucleus, a faint nuclear stellar cluster                    \nl
NGC  6500  &      C   &  inout   &  no     &    Several nuclear sources?              \nl
NGC  6503  &      CS  &  in      &  C     &        A nuclear bar structure, several central sources?             \nl
NGC  6703  &      N   &  --      &  no?    &   No dust                   \nl
NGC  6951  &      LW  &  inout   &  no?    & A star forming ring, dust asociated to the ring with spiral lanes toward the center, several nuclear sources?     \nl
NGC  7177  &      C   &  inout   &  no     &    Very perturbed morphology, chaotic dust through the nucleus             \nl
NGC  7217  &      LW  &  inout   &  no     &    Coherent spiral dust structure, small dust lane through the nucleus                  \nl
NGC  7331  &      C   &  inout   &  Y?     &  Chaotic dust lanes, a nuclear source?                   \nl
NGC  7626  &      N?  &  --      &  no     &     Core light profile. Only a  tiny dust disk or  dust lane through the nucleus                \nl
NGC  7742  &      LW  &  out     &  Y?     & Star forming ring, dust lanes associated to the ring, maybe a compact   nucleus           \nl
\hline
\enddata
\label{tab:dustmorphnucleus}
\end{deluxetable}


\begin{deluxetable}{lcccccccc}
\tabletypesize{\tiny}
\rotate
\tablewidth{0pc}
\tablecaption{Frequency of Dust Morphologies}
\tablehead{
  \colhead{Dust morphology}   &
  \colhead{LLAGNs} &
  \colhead{strong-[OI]}   &
  \colhead{weak-[OI]}        &
   \colhead{Seyferts}    &
  \colhead{Late LLAGNs}        &
  \colhead{Early LLAGNs}        &  
  \colhead{Early type galaxies}   &
  \colhead{Early type (e.l.)}  
  \\
  }
\startdata
No dust	 (N)			&   12\%	&	25\%		&    5\%	&   3 \%    	&  3\%	&  24\%		&     45\%			& 10\%	               \nl
Dust disk (D)			&    9\%	&	15\%		&    5\%	&   0 \%	&  0\%	&  20\%         	&      22\%	 		& 31\%                 \nl 
Chaotic dust (C+CS)	         &   42\%	&	35\%		&   46\%	&   30\%   	&  50\%	& 32\%		&    17\%+7\%		& 35\%                 \nl
Spiral dust (GD+TW+LW)	&   37\%	&	25\%		&   43\%	&   67\%    &  47\%	& 24\%		&    9\%	    		& 14\%                 \nl
\hline
\enddata
\label{tab:observations}
\tablecomments{Col.\ (2-4): this work for LLAGNs, strong-[OI], and weak-[OI] LLAGNs, respectively.
Col.\ (5): Seyfert. galaxies from Martini et al. (2003). 
Col.\ (6-7): this work for early and late type LLAGN hosts, respectively.
Col.\ (8 and 9): Early type galaxies and early type galaxies with emission lines from Lauer et al. (2005).}
\end{deluxetable}


\begin{deluxetable}{lrrrrrrrr}
\tabletypesize{\tiny}
\tablewidth{0pc}
\tablecaption{Total central magnitudes and surface brightness at 0.2, 0.5 and 1 arcsec}
\tablehead{
  \colhead{Galaxy}   &
  \colhead{Filter} &
   \colhead{E(B-V)} &
  \colhead{mag02}   &
  \colhead{mag05}        &
  \colhead{mag1}    &
 \colhead{$\mu$02} &
  \colhead{$\mu$05}   &
  \colhead{$\mu$1}       
  \\
  \colhead{Name}     &
  \colhead{}    &
  \colhead{}     &
  \colhead{mag} &
  \colhead{mag}      &
   \colhead{mag}    &
  \colhead{mag/arcsec$^2$}     &
  \colhead{mag/arcsec$^2$} &
  \colhead{mag/arcsec$^2$}   }
\startdata
NGC 266 & F547m  &  0.069  &  18.43  &  16.73  &  15.60  &  16.28  &  16.69  &  17.20   \nl
NGC 315 & F547m  &  0.065  &  19.52  &  17.48  &  15.97  &  17.31  &  17.19  &  17.28   \nl
NGC 315 & F555w  &  0.065  &  19.36  &  17.42  &  15.94  &  17.24  &  17.15  &  17.25  \nl
NGC 404 & F547m  &  0.059  &  15.91  &  15.16  &  14.58  &  14.42  &  15.99  &  16.72  \nl
NGC 404 & F555w  &  0.059  &  15.92  &  15.15  &  14.55  &  14.37  &  15.94  &  16.67   \nl
NGC 841 & F547m  &  0.048  &  18.03  &  16.23  &  15.07  &  15.78  &  16.14  &  16.79   \nl
NGC 1161 & F547m  &  0.22  &  16.97  &  15.46  &  14.61  &  14.72  &  15.69  &  16.46   \nl
NGC 2685 & F555w  &  0.062  &  17.59  &  16.03  &  15.00  &  15.41  &  16.11  &  16.69  \nl
NGC 2787 & F547m  &  0.13  &  16.84  &  15.43  &  14.50  &  14.73  &  15.63  &  16.34   \nl
NGC 2787 & F555w  &  0.13  &  16.81  &  15.41  &  14.48  &  14.70  &  15.61  &  16.31   \nl
NGC 2911 & F547m  &  0.031  &  18.23  &  16.89  &  15.97  &  16.07  &  17.11  &  17.62   \nl
NGC 3166 & F547m  &  0.031  &  17.87  &  15.84  &  14.68  &  15.45  &  15.74  &  16.30   \nl
NGC 3169 & F547m  &  0.031  &  17.54  &  16.03  &  15.14  &  15.29  &  16.10  &  16.95   \nl
NGC 3226 & F547m  &  0.023  &  17.22  &  15.95  &  15.26  &  15.12  &  16.31  &  17.47   \nl
NGC 3245 & F547m  &  0.025  &  17.07  &  15.60  &  14.60  &  14.93  &  15.76  &  16.24   \nl
NGC 3489 & F555w  &  0.017  &  15.47  &  14.27  &  13.50  &  13.47  &  14.63  &  15.53   \nl
NGC 3992 & F547m  &  0.029  &  17.79  &  16.29  &  15.37  &  15.59  &  16.46  &  17.17   \nl
NGC 3998 & F547m  &  0.016  &  15.93  &  14.77  &  13.91  &  14.11  &  15.10  &  15.73   \nl
NGC 4150 & F555w  &  0.018  &  19.25  &  16.46  &  15.09  &  16.69  &  16.05  &  16.58  \nl
NGC 4192 & F547m  &  0.035  &  17.63  &  16.26  &  15.39  &  15.38  &  16.62  &  17.11   \nl
NGC 4192 & F555w  &  0.035  &  17.63  &  16.25  &  15.38  &  15.37  &  16.59  &  17.11   \nl
NGC 4203 & F547m  &  0.012  &  16.94  &  15.52  &  14.53  &  14.87  &  15.66  &  16.22   \nl
NGC 4203 & F555w  &  0.012  &  16.82  &  15.45  &  14.49  &  14.84  &  15.64  &  16.21   \nl
NGC 4261 & F547m  &  0.018  &  19.13  &  16.94  &  15.32  &  16.73  &  16.59  &  16.54  \nl
NGC 4321 & F555w  &  0.026  &  17.71  &  16.62  &  15.76  &  15.83  &  16.98  &  17.56   \nl
NGC 4450 & F555w  &  0.028  &  17.09  &  15.70  &  14.81  &  15.01  &  15.94  &  16.66   \nl
NGC 4459 & F555w  &  0.046  &  16.62  &  15.16  &  14.29  &  14.49  &  15.35  &  16.15  \nl
NGC 4569 & F547m &   0.046  &  14.43  &  13.76  &  13.29  &  12.85  &  14.65  &  15.78   \nl 
NGC 4736 & F555w  &  0.018  &  15.53  &  14.11  &  13.17  &  13.39  &  14.33  &  14.96  \nl
NGC 4826 & F547m  &  0.041  &  16.50  &  14.94  &  14.00  &  14.23  &  15.09  &  15.84   \nl
NGC 5005 & F547m  &  0.014  &  17.30  &  15.51  &  14.40  &  14.94  &  15.46  &  16.06  \nl
NGC 5055 & F547m  &  0.018  &  15.70  &  14.59  &  14.00  &  13.71  &  15.10  &  16.45   \nl
NGC 5982 & F555w  &  0.018  &  17.80  &  15.89  &  14.81  &  15.39  &  15.86  &  16.51   \nl
NGC 6500 & F547m &   0.090  &  17.97  &  16.32  &  15.36  &  15.60  &  16.45  &  17.18   \nl
NGC 6703 & F555w  &  0.089  &  16.73  &  15.45  &  14.67  &  14.61  &  15.81  &  16.63   \nl
NGC 6951 & F547m  &  0.37  &  17.36  &  16.07  &  15.12  &  15.25  &  16.38  &  16.76   \nl
NGC 7217 & F547m  &  0.088  &  18.05  &  16.14  &  14.98  &  15.71  &  16.06  &  16.60   \nl
NGC 7626 & F555w  &  0.072  &  18.16  &  16.27  &  15.21  &  15.75  &  16.26  &  16.91   \nl
NGC 7742 & F555w  &  0.055  &  16.88  &  15.65  &  14.91  &  14.75  &  15.93  &  16.92   \nl
\hline         
NGC 428 & F606w  &  0.028  &  19.35  &  18.71  &  18.04  &  17.42  &  19.55  &  19.97   \nl
NGC 660 & F606w  &  0.065  &  20.09  &  18.27  &  16.93  &  17.87  &  18.09  &  18.29  \nl
NGC 3368 & F606w  &  0.025  &  16.95  &  15.51  &  14.52  &  14.75  &  15.63  &  16.20   \nl
NGC 3507 & F606w  &  0.024  &  17.46  &  16.51  &  15.75  &  15.53  &  17.05  &  17.62   \nl
NGC 3627 & F606w  &  0.032  &  17.60  &  15.75  &  14.59  &  15.23  &  15.76  &  16.08   \nl
NGC 3675 & F606w  &  0.020  &  17.17  &  15.77  &  14.93  &  15.00  &  15.95  &  16.84   \nl
NGC 3705 & F606w  &  0.046  &  17.63  &  16.21  &  15.46  &  15.48  &  16.34  &  17.55   \nl
NGC 4143 & F606w  &  0.013  &  16.97  &  15.38  &  14.39  &  14.76  &  15.48  &  16.13   \nl
NGC 4314 & F606w  &  0.025  &  17.95  &  16.35  &  15.34  &  15.72  &  16.43  &  17.03   \nl
NGC 4429 & F606w  &  0.033  &  17.24  &  15.94  &  15.07  &  15.14  &  16.28  &  16.84   \nl
NGC 4596 & F606w  &  0.022  &  17.53  &  16.13  &  15.15  &  15.48  &  16.29  &  16.82   \nl
NGC 5005 & F606w  &  0.014  &  17.19  &  15.36  &  14.26  &  14.81  &  15.27  &  15.92   \nl
NGC 5055 & F606w  &  0.018  &  15.54  &  14.49  &  13.91  &  13.56  &  15.05  &  16.33   \nl
NGC 5377 & F606w  &  0.016  &  17.01  &  16.00  &  15.28  &  14.88  &  16.65  &  17.17   \nl
NGC 5678 & F606w  &  0.011  &  18.63  &  17.36  &  16.41  &  16.58  &  17.61  &  18.04   \nl
NGC 5879 & F606w  &  0.012  &  18.95  &  17.33  &  16.18  &  16.76  &  17.38  &  17.67  \nl
NGC 5970 & F606w  &  0.042  &  19.71  &  17.89  &  16.60  &  17.47  &  17.78  &  18.05   \nl
NGC 5985 & F606w  &  0.017  &  18.99  &  17.56  &  16.61  &  16.83  &  17.75  &  18.31   \nl
NGC 6340 & F606w  &  0.049  &  17.46  &  15.92  &  15.05  &  15.25  &  16.14  &  16.94   \nl
NGC 6384 & F606w  &  0.12  &  18.70  &  16.92  &  15.72  &  16.51  &  16.84  &  17.25   \nl
NGC 6503 & F606w  &  0.032  &  18.71  &  17.43  &  16.51  &  16.69  &  17.63  &  18.29   \nl
NGC 6951 & F606w  &  0.37  &  17.26  &  15.94  &  15.02  &  15.10  &  16.24  &  16.68   \nl
NGC 7177 & F606w  &  0.072  &  18.86  &  17.03  &  15.82  &  16.47  &  17.00  &  17.25  \nl
 \hline    
NGC 315 & F814w  &  0.065  &  19.12  &  17.27  &  15.83  &  17.03  &  17.03  &  17.18    \nl
NGC 404 & F814w  &  0.059  &  16.19  &  15.41  &  14.79  &  14.46  &  16.17  &  16.90   \nl
NGC 428 & F814w  &  0.028  &  19.52  &  18.97  &  18.31  &  18.05  &  19.88  &  20.27   \nl
NGC 660 & F814w  &  0.065  &  19.67  &  17.79  &  16.38  &  17.45  &  17.59  &  17.69   \nl
NGC 2685 & F814w  &  0.062  &  17.46  &  15.98  &  15.00  &  15.29  &  16.13  &  16.72   \nl
NGC 2787 & F814w  &  0.13  &  16.80  &  15.37  &  14.44  &  14.63  &  15.57  &  16.23   \nl
NGC 3169 & F814w  &  0.031  &  17.13  &  15.62  &  14.70  &  14.94  &  15.69  &  16.48   \nl
NGC 3368 & F814w  &  0.025  &  16.85  &  15.46  &  14.46  &  14.68  &  15.59  &  16.12   \nl
NGC 3489 & F814w  &  0.017  &  15.80  &  14.52  &  13.70  &  13.71  &  14.83  &  15.63   \nl
NGC 3705 & F814w  &  0.046  &  17.55  &  16.10  &  15.35  &  15.39  &  16.23  &  17.47   \nl
NGC 3998 & F791w  &  0.016  &  15.93  &  14.71  &  13.85  &  13.97  &  15.02  &  15.67   \nl
NGC 4150 & F814w  &  0.018  &  18.48  &  16.20  &  15.05  &  15.99  &  16.04  &  16.70   \nl
NGC 4192 & F791w  &  0.035  &  17.25  &  15.82  &  14.93  &  15.01  &  16.10  &  16.68   \nl
NGC 4192 & F814w  &  0.035  &  17.24  &  15.81  &  14.92  &  15.00  &  16.10  &  16.66   \nl
NGC 4203 & F814w  &  0.012  &  16.85  &  15.40  &  14.41  &  14.72  &  15.54  &  16.14  \nl
NGC 4261 & F791w  &  0.018  &  18.90  &  16.71  &  15.19  &  16.49  &  16.42  &  16.47   \nl
NGC 4314 & F814w  &  0.025  &  17.93  &  16.32  &  15.31  &  15.68  &  16.41  &  17.02   \nl
NGC 4435 & F814w  &  0.030  &  17.69  &  16.00  &  14.93  &  15.37  &  16.04  &  16.53   \nl
NGC 4438 & F814w  &  0.028  &  17.48  &  15.91  &  14.94  &  15.27  &  15.99  &  16.65   \nl
NGC 4450 & F814w  &  0.028  &  17.10  &  15.69  &  14.81  &  14.95  &  15.92  &  16.65   \nl
NGC 4459 & F814w  &  0.046  &  16.58  &  15.09  &  14.23  &  14.39  &  15.28  &  16.14   \nl
NGC 4569 & F791w  &  0.046  &  15.06  &  14.25  &  13.68  &  13.25  &  14.98  &  15.94   \nl
NGC 5005 & F791w  &  0.014  &  17.27  &  15.27  &  14.17  &  14.86  &  15.11  &  15.81   \nl
NGC 5055 & F814w  &  0.018  &  17.00  &  15.13  &  14.15  &  14.78  &  14.98  &  16.08   \nl
NGC 5879 & F814w  &  0.012  &  18.88  &  17.25  &  16.12  &  16.68  &  17.28  &  17.64   \nl
NGC 5982 & F814w  &  0.018  &  17.82  &  15.91  &  14.83  &  15.41  &  15.87  &  16.52   \nl
NGC 6703 & F814w  &  0.089  &  16.73  &  15.45  &  14.67  &  14.57  &  15.82  &  16.63   \nl
NGC 6951 & F814w  &  0.37   &  17.41  &  16.05  &  15.06  &  15.25  &  16.21  &  16.68   \nl
NGC 7177 & F814w  &  0.072  &  18.63  &  16.91  &  15.69  &  16.41  &  16.83  &  17.11   \nl
NGC 7217 & F814w  &  0.088  &  17.71  &  16.00  &  14.88  &  15.52  &  15.91  &  16.50   \nl
NGC 7331 & F814w  &  0.091  &  16.14  &  14.75  &  13.88  &  13.97  &  15.03  &  15.71   \nl
NGC 7626 & F814w  &  0.072  &  18.03  &  16.20  &  15.16  &  15.65  &  16.21  &  16.87   \nl
NGC 7742 & F814w  &  0.055  &  17.14  &  15.80  &  15.03  &  14.96  &  16.03  &  17.00   \nl
\hline
\enddata
\label{tab:observations}
\tablecomments{STSMAG magnitudes (in the F547M, F555W, F606W,  and F814W or F791W) calculated integrating the total flux in a circular aperture of 0.2, 0.5 and 1 arcsec radius (columns 4, 5, and 6, respectively). The surface brightness is calculated at 0.2, 0.5 and 1 arcsec distance from the center  (columns 7, 8 and 9, respectively). }
\end{deluxetable}


\begin{deluxetable}{lccccccc}
\tabletypesize{\tiny}
\tablewidth{0pc}
\tablecaption{Mean values of the central magnitude and surface brightness}
\tablehead{
  \colhead{}   &
  \colhead{mag02} &
  \colhead{mag05}   &
  \colhead{mag1}        &
  \colhead{$\mu$02}    &
 \colhead{$\mu$05} &
  \colhead{$\mu$1}   &
  \colhead{in/out slope}  
  \\
  }
\startdata
All in V				&	17.6	&	16.1		&   15.1	&   15.5	& 16.3		&    16.9		& 2.3	                 \nl
TO					&   17.6	&	16.1		&   15.2	&   15.5	& 16.4		&    17.0	 	& 2.5                \nl 
LINERs				&   17.6	&	16.1		&   15.0	&   15.5	& 16.1		&    16.8		& 2.1                \nl
Young SP				&   17.2	&	15.9		&   14.9	&   15.1	& 16.1		&    16.8	    	& 2.2        \nl
Old SP				&   17.9	&	16.3		&   15.3	&   15.7	& 16.4		&    17.0	    	& 2.1        \nl
\hline
All in I				&	17.4	&	15.9		&    15.1	&   15.2	& 16.0		&    16.6		& 2.0	                 \nl
TO					&   17.5	&	16.0		&    15.1	&   15.3	& 16.2		&    16.8	 	& 2.1                \nl 
LINERs				&   17.4	&	15.8		&    14.7	&   15.2	& 15.8		&    16.4 		& 1.8                \nl
Young SP				&   17.2	&	15.7		&    14.8	&   15.1	& 15.9		&    16.7	    	& 2.2       \nl
Old SP				&   17.7	&	16.1		&    15.0	&   15.4	& 16.1		&    16.7	    	& 1.9        \nl
\hline
\enddata
\label{}
\tablecomments{Magnitudes are tabulated for the central 0.2, 0.5 and 1 arcsec radius (columns 2, 3 and 4), and the surface brightness at  
0.2, 0.5 and 1 arcsec distance from the center (columns 5, 6,  and 7).}
\end{deluxetable}


\begin{deluxetable}{lccccccc}
\tabletypesize{\tiny}
\tablewidth{0pc}
\tablecaption{Frequency of compact nuclear sources in LLAGNs}
\tablehead{
  \colhead{}   &
  \colhead{LLAGNs} &
  \colhead{strong-[OI]}   &
  \colhead{weak-[OI]}        &
  \colhead{YTO}    &
 \colhead{OT} &
  \colhead{YL}   &
  \colhead{OL}  
  \\
    \colhead{}   &
  \colhead{(57)} &
  \colhead{(20)}   &
  \colhead{(37)}        &
  \colhead{(18)}    &
 \colhead{(19)} &
  \colhead{(2)}   &
  \colhead{(18)}  
  \\
  }
\startdata
No-nucleated	 	&	40\%		&	40\%		&  41 \%		&    22\%	&     58\%		&    0\%		&     44\%	               \nl
Nucleated			&      51\%		&	50\%		&   51 \%		&    67\%	&     37\%		&    0\%	 	&     55\%                 \nl 
Clusters+dust?		&       9\%		&	10\%		&    8 \%		&   11\%	&      5\%		&    100\%		&       0\%                 \nl
\hline
\enddata
\label{tab:observations}
\tablecomments{The numbers in brackets indicate the number of objects of each type}
\end{deluxetable}


\begin{deluxetable}{lccccccccc}
\tabletypesize{\tiny}
\tablewidth{0pc}
\tablecaption{Nuclear magnitudes of the nucleated LLAGNs}
\tablehead{
  \colhead{Galaxy Name}   &
  \colhead{Filter} & 
  \colhead{radius}   &
  \colhead{mag}        &
  \colhead{mag}    &
 \colhead{mag} &  
  \colhead{mag}   &
  \colhead{mag}  &
 \colhead{$\mu$ (mag/arcsec$^2$)}&
\colhead{F(15~GHz) (mJy)}
  }
\startdata
NGC 266  &  F547m &     0.14 &  19.72 & 19.59 & 20.08 & 19.63 & 19.84 & 16.65 & 4.1 \nl
NGC 315  &  F555w &     0.14 &  20.70 & 20.64 & 20.96 & 20.73 & 20.73 & 17.62 & 470.0 \nl
NGC 404  &  F555w &     0.23 &  16.03 & 15.98 & 16.12 & 15.98 & 16.11 & 14.06 & $<$1.3 \nl
NGC 428  &  F606w &     0.40 &  19.16 & 19.14 & 19.23 & 19.13 & 19.18 & 18.41 & $<$0.9 \nl
NGC 2787 &  F555w &     0.23 &  17.25 & 17.16 & 17.42 & 17.15 & 17.39 & 15.28 & 7.0 \nl
NGC 3226 &  F547m &     0.23 &  17.53 & 17.44 & 17.71 & 17.43 & 17.68 & 15.56 & 5.4 \nl
NGC 3245 &  F547m &     0.23 &  17.61 & 17.51 & 17.77 & 17.50 & 17.76 & 15.64 & $<$1.0 \nl
NGC 3368 &  F606w &     0.50 &  16.31 & 16.23 & 16.47 & 16.21 & 16.44 & 16.05 & $<$1.0 \nl
NGC 3489 &  F555w &     0.23 &  15.75 & 15.68 & 15.90 & 15.67 & 15.87 & 13.79 & $<$1.0 \nl
NGC 3507 &  F606w &     0.18 &  17.79 & 17.68 & 18.04 & 17.72 & 17.91 & 15.33 & $<$1.5 \nl
NGC 3998 &  F547m &     0.18 &  16.39 & 16.31 & 16.55 & 16.33 & 16.49 & 13.94 & 57.0 \nl
NGC 4203 &  F555w &     0.18 &  17.48 & 17.39 & 17.66 & 17.41 & 17.58 & 15.03 & 9.0 \nl
NGC 4261 &  F547m &     0.14 &  22.32 & 22.22 & 22.52 & 22.35 & 22.36 & 19.24 & 300.0 \nl
NGC 4321 &  F555w &     0.23 &  17.91 & 17.84 & 18.05 & 17.86 & 18.01 & 15.95 & $<$0.9 \nl
NGC 4429 &  F606w &     0.18 &  17.87 & 17.73 & 18.13 & 17.75 & 18.05 & 15.42 & $<$1.1 \nl
NGC 4450 &  F555w &     0.18 &  17.80 & 17.68 & 18.02 & 17.69 & 17.96 & 15.35 & 2.7 \nl
NGC 4459 &  F555w &     0.18 &  17.40 & 17.26 & 17.64 & 17.29 & 17.57 & 14.95 & $<$1.0 \nl
NGC 4569 &  F547m &     0.23 &  14.49 & 14.45 & 14.59 & 14.46 & 14.56 & 12.53 & $<$1.1 \nl
NGC 4596 &  F606w &     0.23 &  17.97 & 17.89 & 18.11 & 17.89 & 18.08 & 16.00 & $<$1.1 \nl
NGC 4736 &  F555w &     0.18 &  16.28 & 16.14 & 16.54 & 16.17 & 16.46 & 13.83 & 1.7 \nl
NGC 5055 &  F547m &     0.27 &  15.74 & 15.67 & 15.84 & 15.66 & 15.86 & 14.17 & $<$1.1 \nl
NGC 5377 &  F606w &     0.32 &  16.75 & 16.72 & 16.83 & 16.71 & 16.82 & 15.52 & 3.1 \nl
NGC 5678 &  F606w &     0.18 &  19.21 & 19.08 & 19.45 & 19.10 & 19.37 & 16.75 & $<$1.0 \nl
NGC 5879 &  F606w &     0.18 &  19.94 & 19.83 & 20.19 & 19.84 & 20.08 & 17.49 & $<$1.1 \nl
NGC 5970 &  F606w &     0.18 &  21.08 & 20.99 & 21.27 & 20.97 & 21.19 & 18.63 & \nodata \nl
NGC 5985 &  F606w &     0.18 &  19.73 & 19.59 & 20.02 & 19.61 & 19.90 & 17.28 & \nodata \nl
NGC 6384 &  F606w &     0.18 &  19.94 & 19.86 & 20.12 & 19.85 & 20.02 & 17.49 & $<$1.0 \nl
NGC 7742 &  F555w &     0.50 &  16.12 & 16.03 & 16.28 & 16.03 & 16.25 & 15.85 & $<$1.1 \nl
\hline
\enddata
\label{tab:observations}
\tablecomments{Column (3): number of arcsec used as radius (r$_{in}$) to measure the nuclear magnitude. 
Column (4-6):  nuclear magnitudes measured within r$_{in}$ , r$_{in}$+1 pixel, r$_{in}$-1 pixel and subtracted 
of  the underlying galaxy contribution assuming that the underlying galaxy have central constant  surface brightness 
that is equal to the value  measured in an annulus with inner and outer radii of r$_{in}$+1 and r$_{in}$+3. 
Column (7-8):  as in column 4 but now the surface brightness of the underlying light contribution is calculated 
at the annulus  with inner and outer radii of r$_{in}$+2 and r$_{in}$+4 (column 7) and inner and outer radii of 
r$_{in}$ and r$_{in}$+2 (column 8). Column (9): nuclear surface brightness estimated using columns (3) and (4).
Column (10): Nuclear radio 15~GHz fluxes from Nagar et al. (2005), upper limits correspond to 5$\sigma$.}
\end{deluxetable}



\clearpage

\begin{figure}[htbp]
\includegraphics{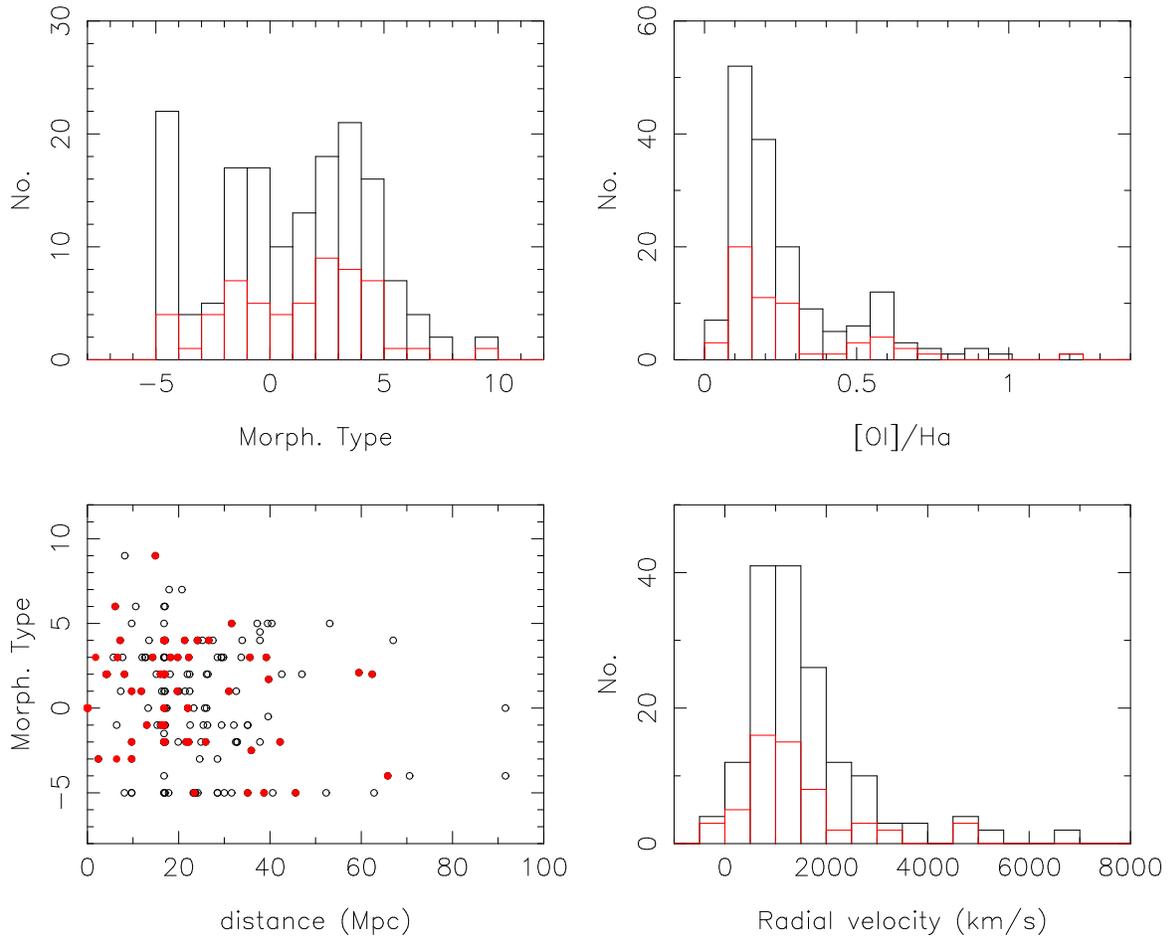}
\vspace{13cm}
\caption{Distances and morphological T types for LLAGN in the HFS97 sample. 
Filled symbols and red lines in the histograms indicate objects in our sample. 
(All data extracted from HFS97.)}
\label{fig:f1}
\end{figure}                                           

\begin{figure}[htbp]
\includegraphics{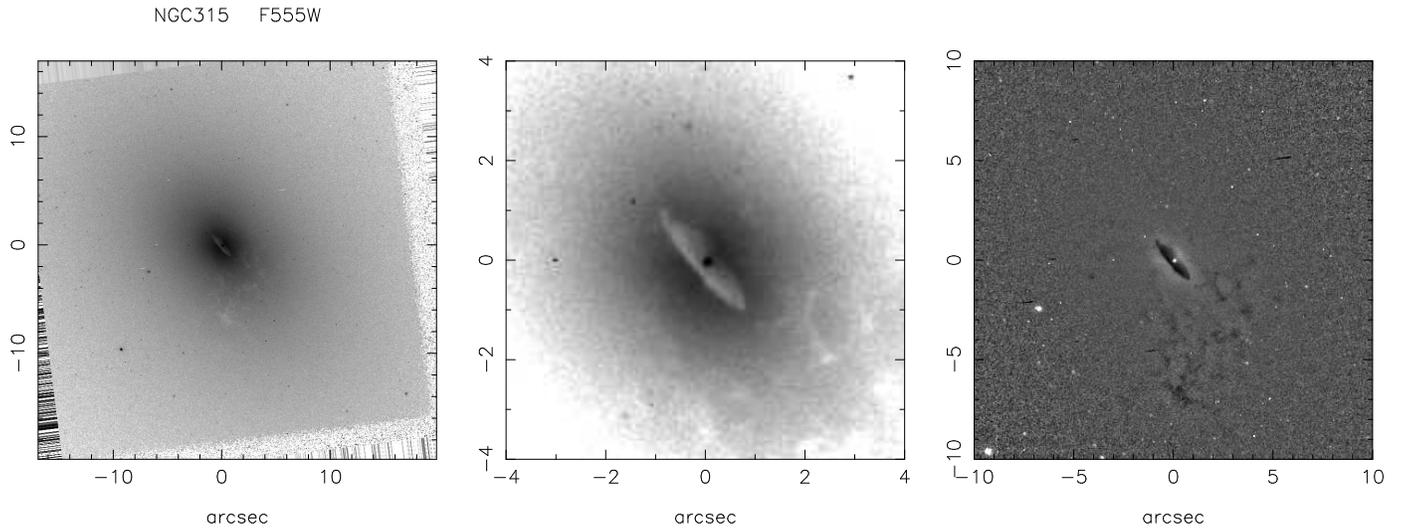}
\vspace{14cm}
\epsscale{0.2}
\caption{Atlas of the HST+WFPC2 images of the sample galaxies at three scales, 
(left:) the full PC frame (or the central $\pm$50 arcsec of the WFPC2 when the center is not in the PC chip; 
(center:) a zoom into the central $\pm$4 arcsec; (right:) the central $\pm$10 arcsec of the median filtered contrast image. }
\label{}
\end{figure}                                           

\begin{figure}[htbp]
\includegraphics{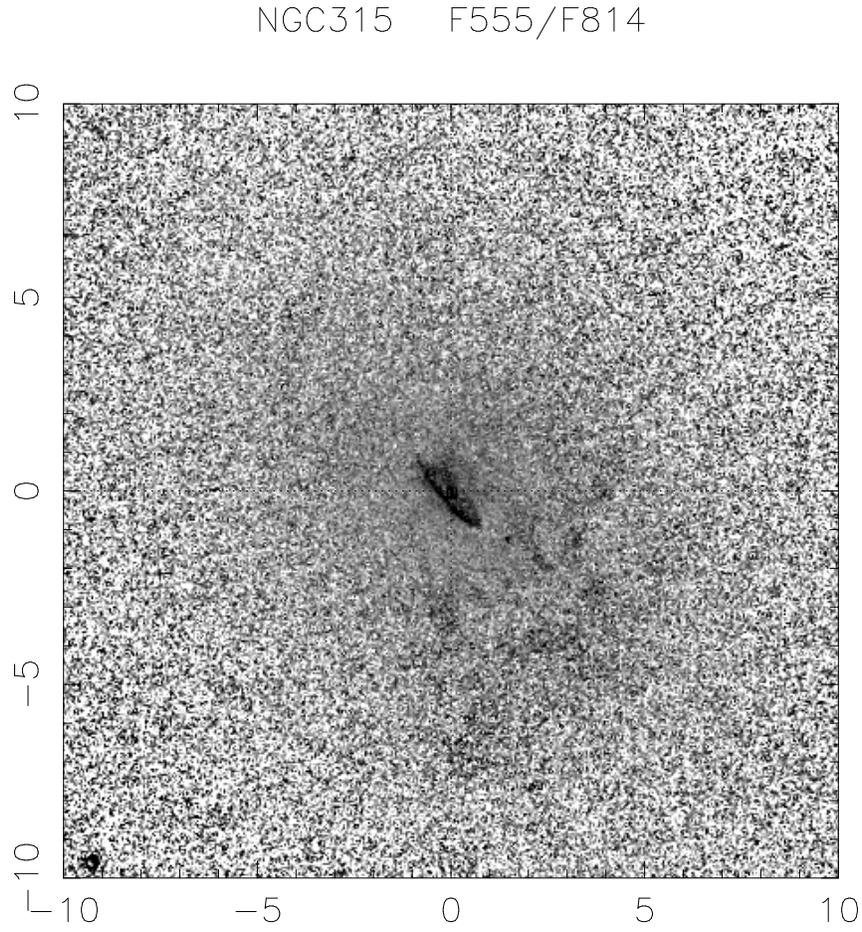}
\vspace{13cm}
\caption{Color images of the galaxies observed in V and I filters. The dotted lines indicate the image center. 
Two sets of this color figures are produced according to the V filter available: (a) F555W/F814W, and (b) F606W/F814W.}
\label{}
\end{figure}                                           

\begin{figure}[ht]
\includegraphics{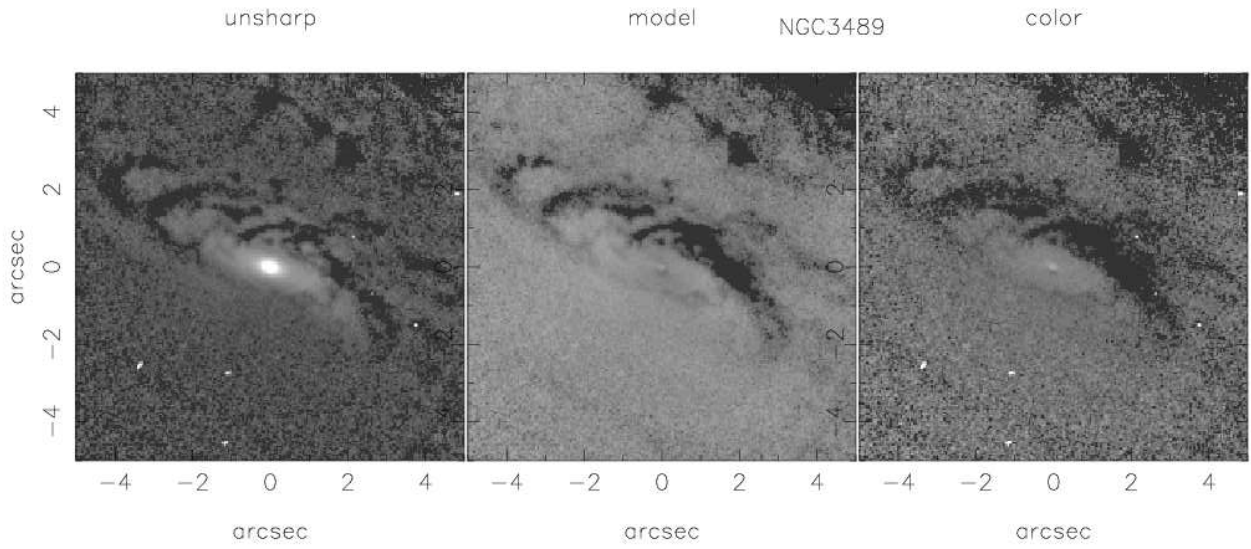}
\vspace{8cm}
\caption{Three images of NGC 3489; (left:) un-sharp image; (center:) original image divided by the isophotal model image model; 
(right:) color image obtained dividing the F555W image by the F814W image.}
\label{fig:f4}
\end{figure}                                           

\begin{figure}[ht]
\includegraphics{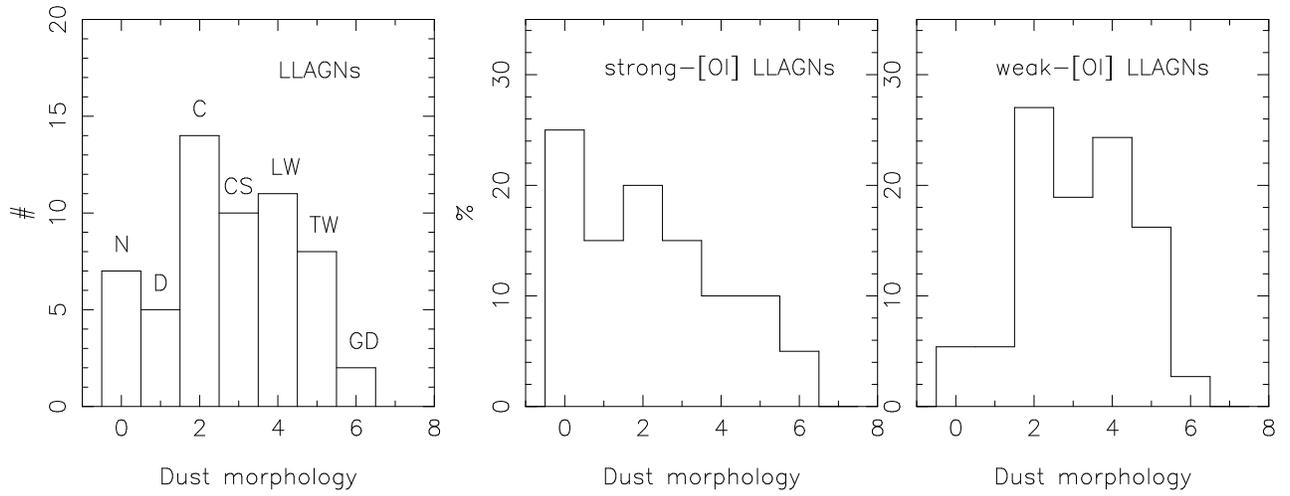}
\vspace{13cm}
\caption{Histogram of the dust morphology distribution: (a), strong-[OI] (b) and weak-[OI] LLAGNs (c), into the seven classes
defined: (0) No dust (N). (1) Disk dust (D). (2) Chaotic circumnuclear dust
(C). (3) Chaotic nuclear spirals (CS). (4) Loosely Wound nuclear spirals
(LW). (5) Tightly Wound nuclear spirals (TW). (6) Grand designed nuclear
spirals (GD).  }
\label{fig:f5}
\end{figure}                                        

\begin{figure}[ht]
\includegraphics{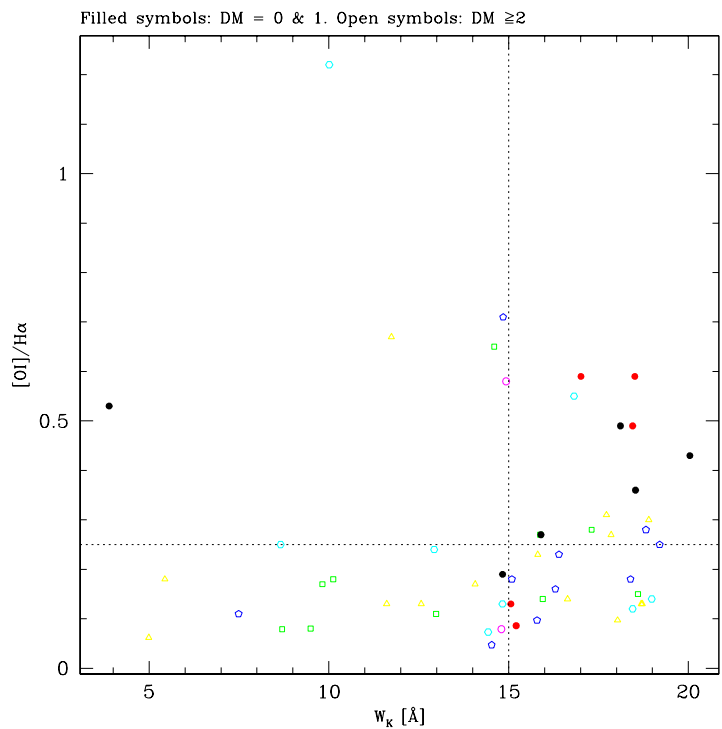}
\vspace{10cm}

\caption{The [OI]/H$\alpha$ ratio vs. equivalent width of the CaII K line, W$_K$. 
Filled symbols are the LLAGNs with dust morphology belonging to the 'no-dust' 
or 'dusty-disk' morphologies, open symbols represent LLAGNs with any other type dust morphology.
No Young LLAGN falls in either the 'no dust' or 'dusty-disk' classes, but old systems 
span the whole range of dust morphologies. Note that the only filled symbol with W$_K\leq$ 15 \AA\ correspond
 to NGC 3998 that is an Old LINER.
 }
\label{fig:f6}
\end{figure}                                           

\begin{figure}[ht]
\includegraphics{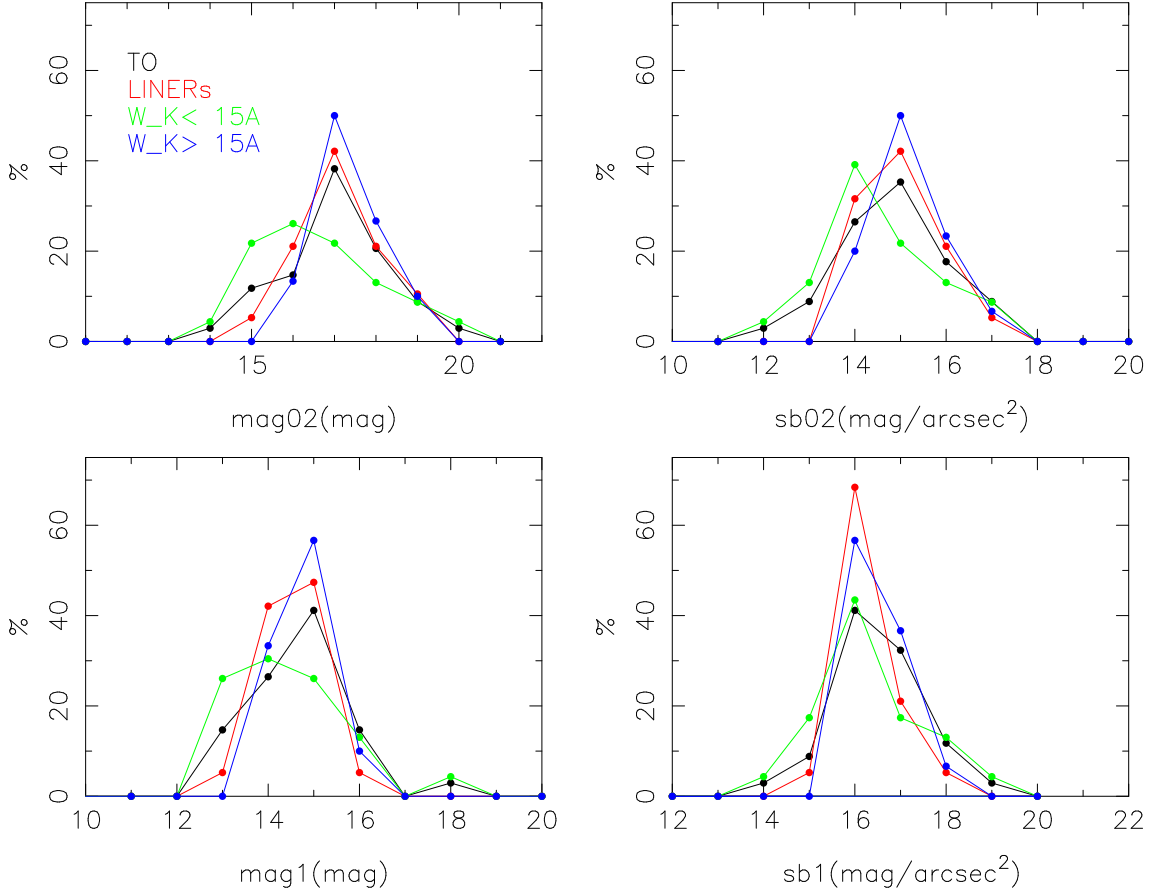}
\vspace{13cm}
\caption{Distribution of the magnitude and the surface
brightness in the F547M, F555W, or F606W  bands. These
distributions are plotted for TOs (black), Liners(red),
LLAGNs with W$_K\leq$15\AA\ (green) (they are presumably LLAGNs
with young stellar population) and LLAGNs with W$_K>$15\AA\ (blue).
Note that these magnitudes have not been corrected by internal dust obscuration. 
However, we expect a larger split between the real distributions of LLAGNs
with young stellar population and LLAGNs with W$_K>$15\AA\  if the former sources 
are the dustier as we estimated in Paper III.}
\label{fig:f7}
\end{figure}                                             

\begin{figure}[ht]
\includegraphics{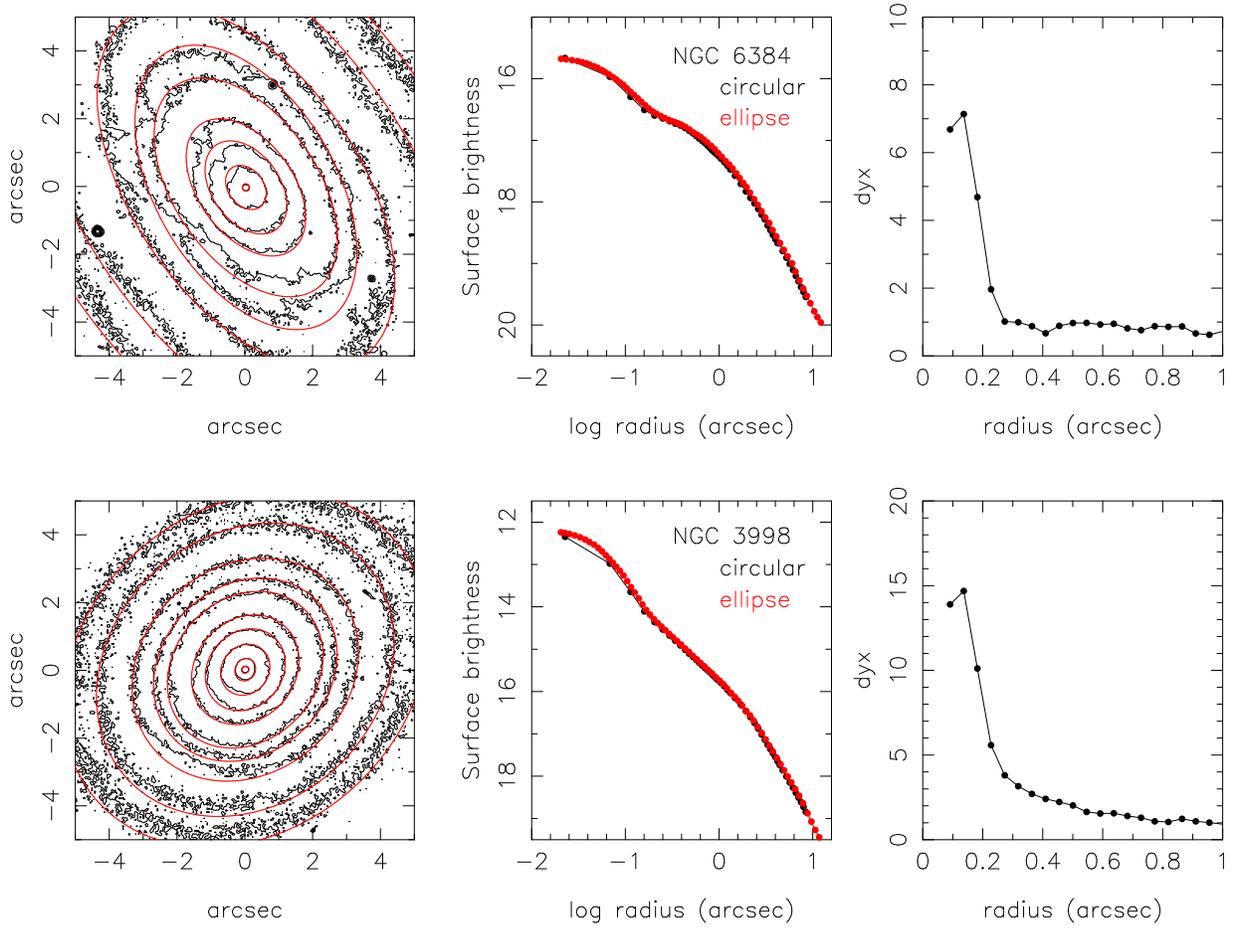}
\vspace{13cm}
\caption{Two examples (NGC 3998 (lower panels) and NGC 6384 (upper panel)  that illustrate the process followed to identify LLAGNs with nuclear compact sources. Left:  The isophotes obtained fitting ellipses (red) to the data (black). Center:  Surface brightness obtained with the circular aperture photometry (black) and fitting ellipses (red). Right: gradient of the surface brightness obtained for the profile derived from the circular aperture photometry.
}
\label{fig:f8}
\end{figure}                                                 

\begin{figure}[ht]
\includegraphics{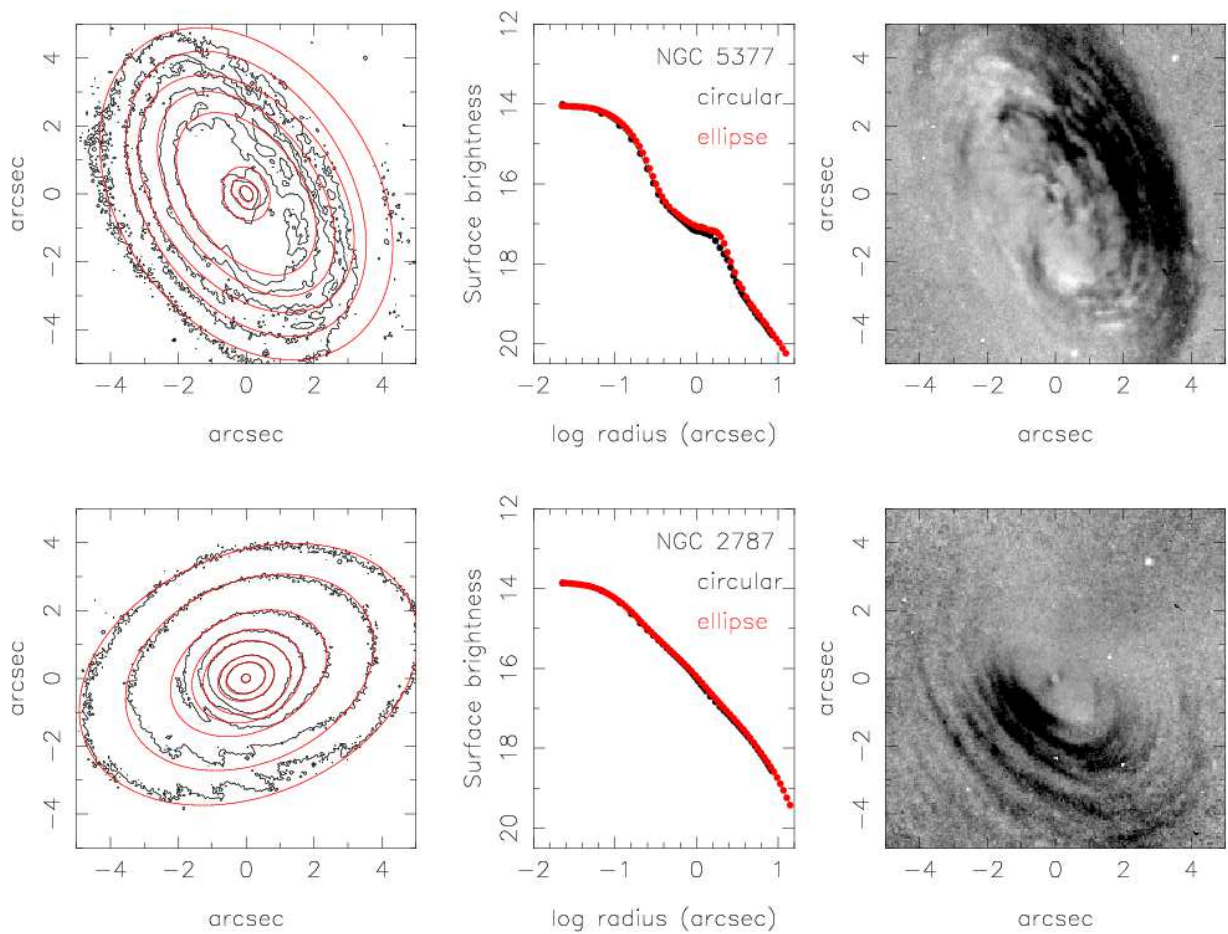}
\vspace{13cm}
\caption{Two examples of the isophotal analysis done (left panels), the surface brightness profiles (center panels), 
and dust obscuration map (right panels). The results are for NGC 5377 (upper panels) and NGC 2787 (lower panels).
Left panels compare the galaxy isophotes (black lines) and the isophotal fits (red line). In the center panels, 
the surface brightness derived from the elliptical isophotal modeling (red points) is compared with the profile obtained 
using aperture photometry (black points). In the dust maps, the obscuration zones are in black. 
}
\label{fig:f9}
\end{figure}                                                              

\begin{figure}[ht]
\includegraphics{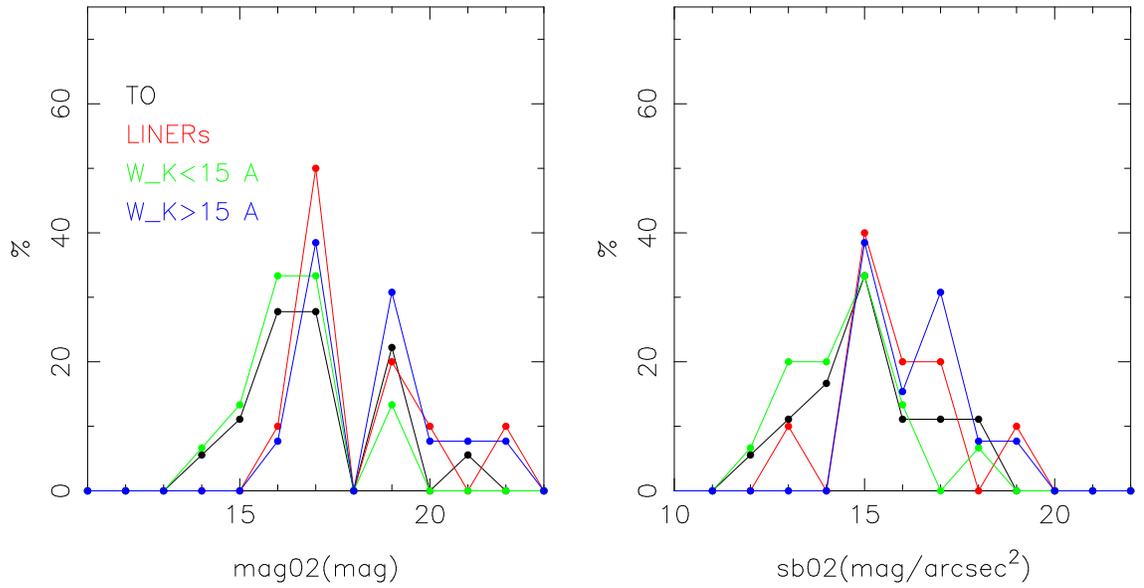}
\vspace{13cm}
\caption{Distribution of the nuclear magnitude in the V band. The
distribution is plotted for TOs (black), Liners(red),
LLAGNs with W$_K\leq$15 \AA\ (green) (they are presumably LLAGNs
with young stellar population) and LLAGNs with W$_K>$15 \AA\ 
(blue).
}
\label{fig:f10}
\end{figure}                                                              

\begin{figure}[htbp]
\includegraphics{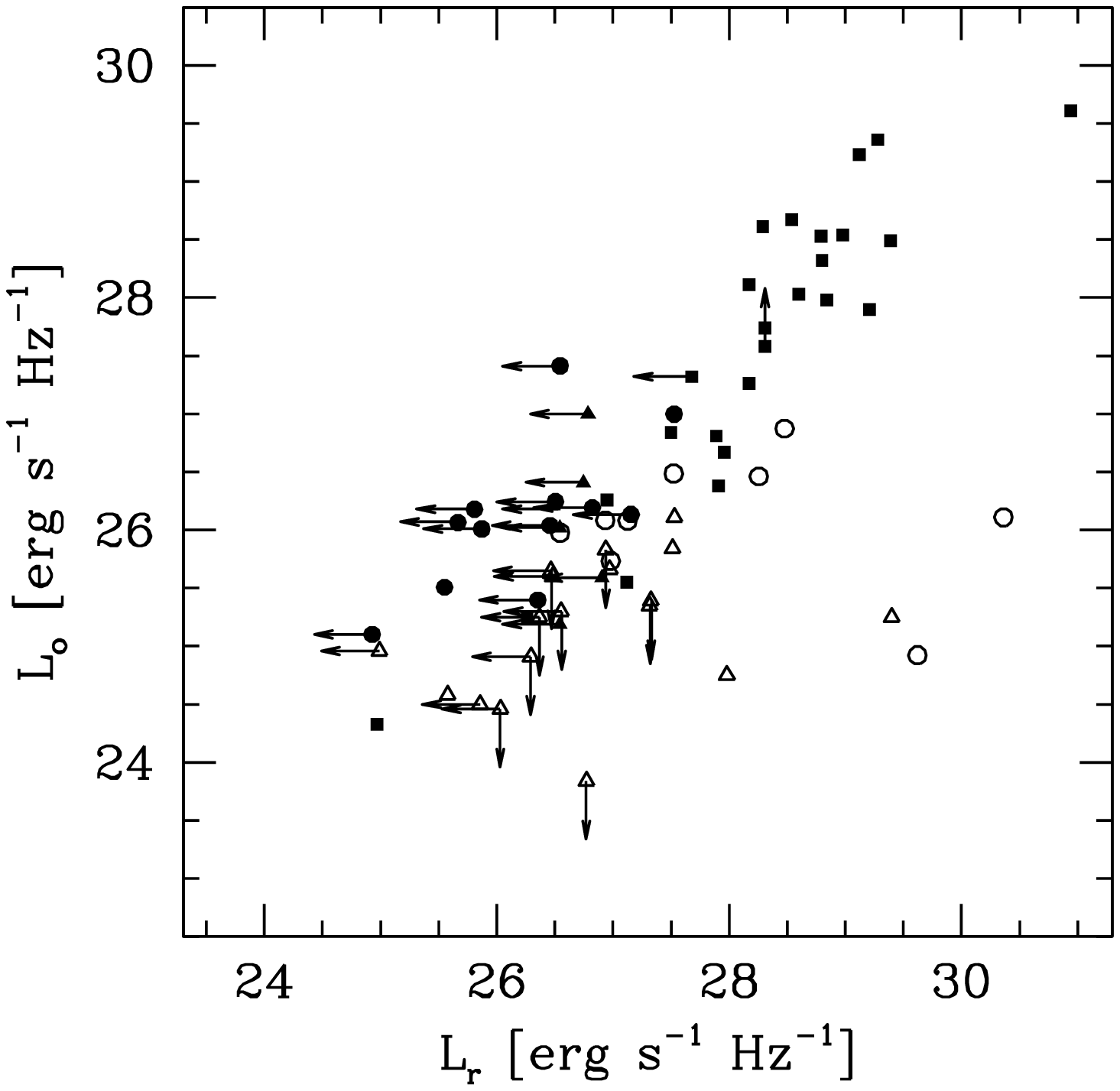}
\vspace{13cm}
\caption{Comparison between the optical (7000 \AA) and radio (5~GHz) luminosities
of Seyfert galaxies (squares) and LINER's (open triangles) from Chiaberge et
al. (2005), with our measurements. Old-LINER's are presented as open circles,
Old-TO's are presented as filled triangles, and Young-TO's are presented
as filled circles.}
\label{fig:f11}
\end{figure}


\begin{references}


\reference{}Binette, L., Magris,ÊC.G., Stasinska,ÊG., \& Bruzual,ÊA.G. 1994, A\&A, 292, 13

\reference{}Boeker,ÊT., Laine,ÊS., vanÊ\,der\,ÊMarel,ÊR.P., Sarzi,ÊM., Rix,ÊH.-W., Ho,ÊL.C., \& Shields,ÊJ.C. 2002, AJ, 123, 1389 

\reference{}Boeker,ÊT., Sarzi,ÊM., McLaughlin,ÊD.E., vanÊ\,der\,ÊMarel,ÊR.P., Rix,ÊH.-W., Ho,ÊL.C., \& Shields,ÊJ.C. 2004, AJ, 127, 105

\reference{}Carollo, M., Stiavelli, M., de\, Zeeuw,ÊP.T., \& Mack, J. 1997, AJ, 114, 2366

\reference{}Carollo, M., Stiavelli, M., de\,ÊZeeuw,ÊP.T., \& Mack,ÊJ. 1998, AJ, 116, 68

\reference{}Carollo, M., Stiavelli, M., Seigar,ÊM., deÊZeeuw,ÊP.T., \& Dejonghe,ÊH. 2002, AJ, 123, 159 2002 

\reference{}Chiaberge, M., Capetti, A., \& Celotti, A. 1999, A\&A, 349, 77

\reference{}Chiaberge, M., Capetti, A., \& Macchetto, F. D. 2005, ApJ, 625, 716

\reference{}Cid Fernandes, R., Heckman, T., Schmitt, H., Gonz\'alez Delgado, R.M., \& Storchi-Bergmann, T. 2001a, ApJ, 558, 81

\reference{}Cid Fernandes, R., Gonz\'alez Delgado, R.M., Schmitt, H., et al. 2004a, 605, 105

\reference{}Cid Fernandes, R., Gu, Q., Melnick,ÊJ., Terlevich,ÊE., Terlevich,ÊR., Kunth,ÊD., RodriguesÊLacerda,ÊR., \& Joguet,ÊB. 2004b, MNRAS, 355, 273 

\reference{}Cid Fernandes, R., Gonz\'alez Delgado, R.M., Storchi-Bergmann,ÊT., Martins,ÊL.P., \& Schmitt,ÊH. 2005, MNRAS, 356, 270

\reference{}Colina, L., Gonz\'alez Delgado, R.M., Mas-Hesse, J.M., \& Leitherer, C. 2002, ApJ, 579, 545

\reference{}C\^ot\'e,ÊP., et al. 2006, ApJS, 165, 57

\reference{}de Vaucouleurs, G., de Vaucouleurs, A., Corwin, H. G., Jr., Buta, R. J., Paturel, G.; Fouque, P., 1991, Third Reference Catalogue of Bright Galaxies (New York: Springer) (RC3)

\reference{}Dudik, R.P., Satyapal, S., Gliozzi, M., \& Sambruna, R.M. 2005, ApJ, 620, 113

\reference{}Ferrarese, L., \& Merrit, D. 2000, ApJ, 539, L9 

\reference{}Ferrarese, L., C\^ot\'e,ÊP., Jord\'an,ÊA., et al. 2006a, ApJS, 164, 334

\reference{}Ferrarese, L., C\^ot\'e,ÊP., DallaÊBont\`a,ÊE., et al. 2006b, ApJ, 644, 21

\reference{}Ferrarese, L., C\^ot\'e,ÊP., Blakeslee, J.P., Mei, S., Merrit, D., West, M.J. 2006c, (astro-ph/0612139)

\reference{}Ferland, G.J., \& Netzer,ÊH. 1983, ApJ, 264, 105

\reference{}Filippenko, A.V., \& Terlevich, R. 1992, AJ, 397, 79

\reference{}Gebhardt, K., Kormendy,ÊJ., \& Ho,ÊL.C. 2000, ApJ, 539, L13 

\reference{}Gonz\'alez Delgado, R.M., Heckman, T., Leitherer, C., Meurer, G., Krolik, J., Wilson, A.S., Kinney, S., \& Koratkar, A. 1998, ApJ, 505, 174

\reference{}Gonz\'alez Delgado, R.M., Heckman, T., \& Leitherer, C. 2001, ApJ, 546, 845

\reference{}Gonz\'alez Delgado, R.M., Cid Fernandes, R.,  P\'erez, E., Martins, L. P., Storchi-Bergmann, T., Schmitt, H., Heckman, T.M.,  \& Leitherer, C. 2004, 605, 127

\reference{}Gonz\'alez-Mart\'\i n, O., Masegosa, J., M\'arquez, I., Guerrero, M.A., \& Dultzin-Hacyan, D. 2006,
A\&A, 460, 45

\reference{}Graham, A.W., Erwin, P., Trujillo, I., \& Asensio, R.A. 2003, AJ, 125, 2951

\reference{}Heckman, T. M. 1980, A\&A, 87, 152

\reference{}Ho, L.C., Filippenko, A.V. \& Sargent, W.L.W. 1995, ApJ, 98, 477

\reference{}Ho, L.C., Filippenko, A.V. \& Sargent, W.L.W. 1997a, ApJS, 112, 315 

\reference{}Ho, L.C., Filippenko, A.V. \& Sargent, W.L.W. 1997b, ApJS, 112, 391

\reference{}Ho,ÊL.C., Rudnick,ÊG., Rix,ÊH.-W., Shields,ÊJ.C., McIntosh,ÊD.H., Filippenko,ÊA.V., Sargent,ÊW.L.W., \&  Eracleous,ÊM.  2000, ApJ, 541, 120

\reference{}Holtzman, J.A., Burrows, C., Casertano, S., Hester, J., Trauger, J.T., Watson, A.M., Worthey, G. 1995, PASP, 107, 1065

\reference{}Hughes, M. A., Axon, D., Atkinson, J., Alonso-Herrero, A., Scarlata, C., Marconi, A., Batcheldor, D., Binney, J.  2005, AJ, 130, 73

\reference{}King, I.R. 1966, AJ, 71, 64

\reference{}Lauer, T.R., et al. 1995, AJ, 110, 2622

\reference{}Lauer, T.R., et al. 2005, AJ, 129, 2138

\reference{}Lauer, T.R., et al. 2006, (astro-ph/0609762) 

\reference{}Maoz, D., Koratkar, A., Shields, J.C., Ho, L.C., Filippenko, A.V., \& Sternberg, A. 1998, AJ, 116, 55

\reference{}Maoz, D., Nagar,ÊN.M., Falcke,ÊH., \& Wilson,ÊA.S. 2005, ApJ, 625, 699

\reference{}Martini, P., \& Pogge, R.W. 1999, AJ, 118, 2646

\reference{}Martini, P., Regan, M.W., Mulchaey, J.S., Pogge, R.W. 2003, ApJ, 589, 774

\reference{}Meurer, G.R., Heckman, T.M., Leitherer,ÊC., Kinney,ÊA., Robert,ÊC., \& Garnett,ÊD.R. 1995, AJ, 110, 2665

\reference{}Mu\~noz-Mar\'\i n, V., Gonz\'alez Delgado, R.M., Schmitt, H., et al. 2007, AJ, 134, 648

\reference{}Nagar, N.M., Falcke,ÊH., Wilson,ÊA.S., \& Ho,ÊL.C. 2000, ApJ, 542, 186

\reference{}Nagar, N.M., Falcke,ÊH., Wilson,ÊA.S., \& Ulvestad,ÊJ.S. 2002, A\&A, 392, 53 

\reference{}Nagar, N. M., Falcke, E., \& Wilson, A. S. 2005, A\&A, 435, 521

\reference{}Pogge, R.W., \& Martini, P. 2002, ApJ, 569, 624

\reference{}Ravindranath, S., Ho, L.C., Peng, C.Y., Filippenko, A.V., Sargent, W.L.W. 2001, AJ, 122, 653

\reference{}Regan, M.W., \& Mulchaey, J.S. 1999, AJ, 117, 2676

\reference{}Rest, A., van den Bosch, F.C., Jaffe, W., Tran, H., Tsvetanov, Z., Ford, H.C., Davies, J., Schafer, J. 2001, AJ, 121, 2431
 
\reference{}Satyapal, S., Sambruna,ÊR.M., \& Dudik,ÊR.P. 2004, A\&A, 414, 825

\reference{}Scarlata, C., Stiavelli, M., Hughes, M. A., Axon, D., Alonso-Herrero, A., Atkinson, J., Batcheldor, D., Binney, J. et al. 2004, AJ, 128, 1124

\reference{}Schlegel, D.J., Finkbeiner, D.P., \& Davis, M. 1998, ApJ, 500, 525

\reference{}S\'ersic, J.-L. 1968, Atlas de Galaxias Australes (C\'ordova: Obs. Astron., Univ. Nac. C\'ordova)

\reference{}Shields, J.C., Rix,ÊH.-W., McIntosh,ÊD.H., Ho,ÊL.C., Rudnick,ÊG., Filippenko,ÊA.V., Sargent,ÊW.L.W., \& Sarzi,ÊM. 2000, ApJ, 534, L27

\reference{}Simoes Lopes, R.D., Storchi-Bergmann, T., de F\'atima Saraiva, M., Martini, P. 2007, ApJ, 655, 718

\reference{}Storchi-Bergmann, T., Eracleous, M., Livio, M., Wilson, A.S., Filippenko, A.V., \& Halpern, J.P. 1995, ApJ, 443, 617

\reference{}Taniguchi, Y., Shioya, Y.,\&  Murayama, T. 2000, AJ, 120, 1265
 
\reference{}Terashima, Y., Ho, L.C., \& Ptak, A.F. 2000, ApJ, 539, 161

\reference{}Tran, H.D., Tsvetanov, Z., Ford, H.C., Davies, J., Jaffe, W., van der Bosch, F.C., \& Rest, A. 2001, AJ, 121, 2928

\reference{}Trujillo, I., Erwin, P., Asensio Ramos, A, \& Graham, A.W. 2004, AJ, 127, 1917

\reference{}Tully, R.B., 1988, Nearby Galaxies Catalog (Cambridge: Cambridge Univ. Press)

\reference{}van der Marel, R. P., Rossa, J., Walcher, C. J., Boeker, T., Ho, L. C., Rix, H.-W., \& Shields, J. C. 2007
in Stellar Populations as Building Blocks of Galaxies,  eds. A. Vazdekis and R. Peletier (astro-ph/0702433)

\reference{}Verdoes Kleijn, G.A., \& de Zeeuw, P.T. 2005, A\&A, 387, 441
 
\end{references}
\end{document}